\documentclass[10pt,journal,compsoc]{IEEEtran}

\usepackage{subfigure}
\usepackage{xspace}
\usepackage{multirow, booktabs}
\usepackage{siunitx}
\usepackage{enumitem}
\usepackage{makecell}
\usepackage{graphicx}
\usepackage{amsmath}
\usepackage{balance}
\usepackage{color}

\def\ie{\textit{i.e.}\xspace}
\def\etc{\textit{etc}\xspace}
\def\eg{\textit{e.g.}\xspace}
\def\etal{\textit{et~al.}\xspace}

\ifCLASSOPTIONcompsoc
  \usepackage[nocompress]{cite}
\else
  \usepackage{cite}
\fi

\newcommand\qi[1]{{\color{black}{{{#1}}}}}
\def\re2#1{{\color{black}#1}}
\def\red1#1{{\color{black}#1}}
\def\reda3#1{{\color{black}#1}}
\begin{document}
%
\title{Securing Face Liveness Detection Using Unforgeable Lip Motion Patterns}

\author{Man~Zhou,
        Qian~Wang,~\IEEEmembership{Senior Member,~IEEE,}
        Qi~Li,~\IEEEmembership{Senior Member,~IEEE,}
        Peipei~Jiang,
        Jingxiao~Yang,
        Chao~Shen,~\IEEEmembership{Senior Member,~IEEE,}
        Cong~Wang,~\IEEEmembership{Fellow,~IEEE,}
        and~Shouhong~Ding

\IEEEcompsocitemizethanks{
\IEEEcompsocthanksitem M. Zhou, Q. Wang and P. Jiang are with
the School of Cyber Science and Engineering, Wuhan University, Wuhan 430072, China (e-mail: \{zhouman, qianwang, ppjiang\}@whu.edu.cn).
\IEEEcompsocthanksitem Q. Li is with the Institute for Network Sciences and Cyberspace, Tsinghua University, Beijing 100084, China (e-mail: qli01@tsinghua.edu.cn).
\IEEEcompsocthanksitem J. Yang is with the School of Computer Science, Wuhan University, Wuhan 430072, China (e-mail: yangjingxiao@whu.edu.cn).
\IEEEcompsocthanksitem C. Shen is with the MOE Key Laboratory for Intelligent Networks and Network Security, Xi'an Jiaotong University, Xi'an 710049, China, and also with the School of Cyber Science and Engineering, Xi'an Jiaotong University, Xi'an 710049, China (e-mail: chaoshen@mail.xjtu.edu.cn).
\IEEEcompsocthanksitem C. Wang is with the Department of Computer Science, City University of Hong Kong, Hong Kong (e-mail: congwang@cityu.edu.hk).
\IEEEcompsocthanksitem S. Ding is with Youtu Lab, Tencent, Shanghai, China (e-mail: ericshding@tencent.com).
}
}

\IEEEtitleabstractindextext{%

\begin{abstract}
Face authentication usually utilizes deep learning models to verify users with high recognition accuracy. However, face authentication systems are vulnerable to various attacks that cheat the models by manipulating the digital counterparts of human faces. So far, lots of liveness detection schemes have been developed to prevent such attacks. Unfortunately, 
the attacker can still bypass these schemes by constructing wide-ranging sophisticated attacks. We study the security of existing face authentication services (\eg, Microsoft, Amazon, and Face++) and typical liveness detection approaches.
Particularly, 
we develop a new type of attack, \ie, the low-cost 3D projection attack that projects manipulated face videos on a 3D face model, which can easily evade these face authentication services and liveness detection approaches.
To this end, we propose FaceLip, a novel liveness detection scheme for face authentication, which utilizes unforgeable lip motion patterns built upon well-designed acoustic signals 
to enable a strong security guarantee. 
\qi{The unique lip motion patterns for each user are unforgeable because FaceLip verifies the patterns by capturing and analyzing the acoustic signals that are dynamically generated according to random challenges, which ensures that our signals for liveness detection cannot be manipulated.
}
Specially, we develop robust 
algorithms for FaceLip to eliminate the impact of noisy signals in the environment 
and thus can accurately infer the lip motions at larger distances. 
We prototype FaceLip on off-the-shelf smartphones and conduct extensive experiments under different settings. Our evaluation with 44 participants
validates the effectiveness and robustness of FaceLip.
\end{abstract}

}

\maketitle

\IEEEdisplaynontitleabstractindextext

\IEEEpeerreviewmaketitle

\IEEEraisesectionheading{\section{Introduction}}
\IEEEPARstart{F}{ace} authentication has already been widely deployed, \eg, in phone unlock~\cite{FaceID}, online payment~\cite{Alipay}, and door entrance~\cite{DoorEntry}, as it provides a simple yet effective recognition and identification method for mobile devices. 
A data analysis report~\cite{FaceMarket} shows that the facial recognition market is expected to reach \$9.06 billion in 2024. 
However, prior studies show that face authentication systems are vulnerable to various attacks and can be exploited because the deep learning algorithms used in these systems are susceptible to well-designed adversarial examples~\cite{yuan2019adversarial}.
Moreover, with the increase in the popularity of social network software, \eg, Facebook, it is easier for an attacker to access photos or facial videos of the victim and construct a fake facial model.
For example, an attacker can launch the media based facial forgery (MFF) attack and easily 
impersonate the target face by leveraging advanced algorithms and high-resolution printers/screens~\cite{FaceFlashing18}.

To defeat these attacks, various liveness detection methods have been developed. 
For instance, challenge-response based approaches require a user to interact with the system, \eg, smiling, blinking eyes, or rotating heads~\cite{pan2007eyeblink}, to detect fake faces. 3D shape based approaches identify fake faces by extracting the 3D structure of a human face from the captured video~\cite{li2015seeing}.  
As for texture pattern based approaches, the attacks are detected by analyzing color and micro-textures~\cite{boulkenafet2016face}.
Unfortunately, these liveness detection approaches can still be easily bypassed by constructing sophisticated 3D dynamic attacks.

\begin{figure}[!t]
\centering
\includegraphics[width=0.9\columnwidth]{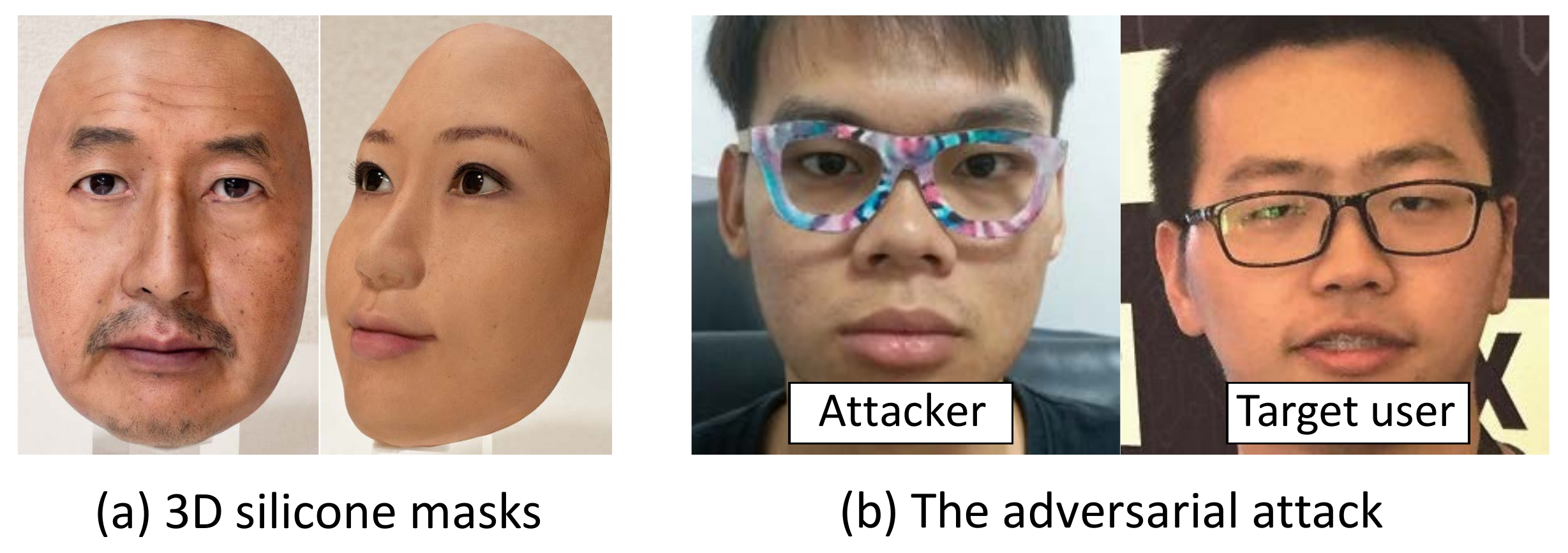}
\vspace{-2mm}
\caption{Examples of 3D dynamic attacks.}
\label{fig:two-attacks}
\vspace{-3mm}
\end{figure}

For instance, these approaches are unable to resist 3D mask attacks~\cite{erdogmus2014spoofing,bhattacharjee2018spoofing,ramachandra2019custom} that leverage high-quality custom 3D silicone masks (see Fig.~\ref{fig:two-attacks} (a)).
Moreover, physical-world adversarial samples can be constructed to realize unobtrusive perturbations (\eg, eyeglass frames) around the attacker's faces~\cite{sharif2016accessorize,zhou2018invisible,Komkov2019AdvHat} and cheat face recognition systems, which is called the adversarial attack, 
as depicted in Fig.~\ref{fig:two-attacks} (b).

\qi{The vulnerabilities are incurred due to the inherent flaw of the liveness detection approaches, \ie, 
the digital counterparts used in these approaches can be faked, replayed, and manipulated. }  
In order to study these issues, we evaluate the security of face authentication systems by evaluating mainstream commercial facial recognition services and typical liveness detection approaches. 
In particular, we develop a new type of 3D dynamic attacks, \ie, the 3D projection attack. It is able to evade all these liveness detection approaches by dynamically reconstructing the media of users under authentication, which is a low-cost but effective attack.
In the attack, we synthesize the victim's face video containing interactions with the system in real time to easily bypass the challenge-response based detection. Particularly, compared to the existing 3D dynamic attacks, the attack is generic and cost-efficient. Our experiments demonstrate that %
our proposed attack
can easily bypass all mainstream commercial face authentication services and liveness detection approaches. Thus, it is vital to have a secure liveness detection method that can ensure the security of face authentication.

\qi{We propose FaceLip, a lip motion based secure face liveness detection scheme that enables a tie-breaking method to defeat existing attacks by utilizing unforgeable lip motion patterns. We leverage the unique lip motion of each individual~\cite{faraj2006motion} to construct an unforgeable pattern. The pattern is unforgeable because we capture the pattern in the form of acoustic signals that are dynamically generated for the user according to random challenges in each round of detection.
FaceLip verifies the pattern by analyzing the captured acoustic signals to perform liveness detection, which cannot be faked, replayed, and manipulated. 
Note that, although several lip-reading based authentication schemes~\cite{lu2019lippass,tan2018silentkey} also utilize acoustic signals. 
They are vulnerable to various attacks, \eg, replay attacks and synthesis attacks
, because they cannot differentiate between fresh signals generated by users and replayed (or synthesized) signals. 
FaceLip leverages the speaker of a smartphone to play imperceptible acoustic signals and utilize the microphone to record the reflected signals from the lips for liveness detection. In particular, by eliminating the impact of environmental noise, FaceLip can accurately capture users' unique lip motions even in larger distance scenarios, \ie, the distance 
is larger than 18cm that is the minimum distance we measured to capture the whole face for authentication. These problems are not well addressed in the literature. Thus, to the best of our knowledge, FaceLip is the first practical method that utilizes unforgeable lip motion patterns built upon well-designed acoustic signals to enable secure face liveness detection.
}


\qi{FaceLip utilizes unforgeable lip motions as patterns for liveness detection, and thus it is robust and secure under wide-ranging attacks. To achieve this, we develop two key modules in FaceLip. 
The \textit{Motion Verification} module detects whether the recorded signals contain ``physical'' lip motions and guarantees the signals are not relayed or manipulated, \eg, defeating fake signals constructed by MFF attacks. In order to ensure the captured lip motions are only produced from the user's ``physical'' movements rather than ``visual'' movements,
we transform the acoustic system of a phone into an active sonar and compute a fine-grained estimation of subtle lip movements by leveraging the phase shift based approach, which can achieve millimeter-level measurement accuracy~\cite{WuYZCW19}.
Especially, FaceLip leverages dynamic and static interference elimination to remove the multi-path interference and obtain the acoustic signal components only related to lip motions, which guarantees that the signals are accurate enough for large distance liveness detection.
The \textit{Consistency Verification} module abstracts the unique lip motion pattern of each individual and verifies if the pattern is the real one of the user, which guarantees that the ``physical'' information of lip motions constructed by sophisticated attacks, \eg, 3D dynamic attacks, cannot bypass our system.
It characterizes the lip motion pattern of each individual by extracting robust energy-band time-frequency features and builds an efficient binary classifier for each user to determine whether each input of liveness detection is consistent with the real user.
}

Our contributions are summarized as follows.
\begin{itemize}[leftmargin=*]
\setlength{\itemsep}{0.2pt}
  \item We study the security of existing commercial facial recognition services and typical liveness detection schemes and find they are vulnerable to a new 3D projection attack.
  \item We propose FaceLip, which enables a tie-breaking mechanism of lip motion based liveness detection by leveraging unforgeable acoustic signals, instead of using traditional image or video signals.
  \item We develop robust algorithms to eliminate the impacts of noisy signals in the environment 
  and accurately infer lip motions for large distance liveness detection so that acoustic signals can be applied in liveness detection for face authentication. 
  \item We prototype FaceLip on off-the-shelf smartphones and conduct extensive experiments to evaluate its effectiveness and robustness. Our evaluation with 44 participants illustrates that it achieves over 95\%  overall detection accuracy at around 5\% EER.

\end{itemize}


\section{Threat Model and Related Work}
\subsection{Threat Model}
Existing face authentication attacks can be categorized into  presentation attacks and compromising attacks~\cite{uzun2018rtcaptcha}, according to how the malicious face media is fed into the system.  Presentation attacks physically ``present'' the malicious face media to the front-end device of the system, while compromising attacks bypass the camera and fabricate the digital output of the front-end device. In general, it is more difficult to construct compromising attacks than presentation attacks, since the former must compromise the integrity of the front-end devices of the system (\eg, hacking the device). Considering that the front-end device is usually protected effectively in the real world, we focus on defeating various sophisticated presentation attacks.

\qi{In this paper, we particularly consider practical attacks that can be constructed in real-world scenarios, \eg, unattended automatic face authentication system. To construct presentation attacks, an attacker could control the environment but cannot manipulate the front-end devices (\eg, cameras). }
The attacker aims to bypass the system by replaying the victim's face video or impersonating the victim. The attacker can get a large number of photos or facial videos of the victim, \eg, from the social networks, to build the facial model of the targeted individual and synthesize the impersonating media (\eg, pictures and videos) of the target face.
We consider a typical presentation attack, \ie, the MFF attack, 
as well as a variety of sophisticated 3D dynamic attacks, \eg, the 3D mask attack~\cite{erdogmus2014spoofing}, the adversarial attack~\cite{sharif2016accessorize}, and the 3D projection attack developed by ourselves.
Note that, we do not consider other types of 3D dynamic attacks, \eg, performing cosmetic surgery or using a person with similar looks, because they are very difficult to launch in real-world settings~\cite{FaceFlashing18}.

For simplicity, in this paper, we use Android phones as an example to show the effectiveness of our proposed liveness detection method. However, the method is generic and can be readily deployed in various terminal devices equipped with a camera, a speaker, and a microphone, \eg, various smartphones (including iPhones) and door entry devices.

\subsection{Related Work}
In this section, 
we classify existing face liveness detection methods 
into three categories: challenge-response based approaches, facial 3D shape based approaches, and texture pattern based approaches. In addition, we discuss acoustic signal based techniques that can be potentially used in face liveness detection.


\noindent{\bf Challenge-Response Based Approaches.} In these approaches, a user is required to interact with the liveness detection device,~\eg, eye-blink~\cite{pan2007eyeblink}, or move the lips~\cite{KollreiderFFB07}. When the device presents a challenge, the user needs to respond accordingly, and the response is evaluated to detect if the user is valid.
Normally, these interactions can effectively resist the still image attacks~\cite{li2014understanding}, as the static photo is not able to make a movement to respond to the challenge. \red1{However, these approaches are vulnerable to dynamic attacks. For example, 
video based attacks can easily bypass them 
since they can use arbitrary facial frames to create a motion that fulfills the desired challenges.} Simple motions like eye-blink simulated by 
displaying different photos~\cite{HowtobypassAndroid} or manufacturing a 3D mask resembled the victim's face~\cite{erdogmus2014spoofing} 
can also circumvent the detection. 
It can mitigate the issue above by 
increasing the randomness of the challenges and performing the time verification of the responses, \eg, by utilizing audio captchas~\cite{gao2010audio}.
The captcha based defenses can defeat automatical attacks but cannot defend against the presentation attacks where a real person is involved in solving the captchas.

\noindent{\bf Facial 3D Shape Based Approaches.}
\red1{This type of approaches detects attacks by distinguishing facial structures recovered from real faces contain stereo characters from the ones recovered from photos/videos that are usually planar in depth.}
There are mainly two types of 3D feature extractions: appearance-based and motion-based approaches.
The former analyzes the visual-spatial data of the face, 
\eg, analyzing the face by using an optoelectronic stereo device~\cite{lagorio2013liveness}.
Moreover, the facial landmark techniques~\cite{wang2013face}, optical flows~\cite{bao2009liveness}, and the deformation of pixels~\cite{kim2013face} can also be used to distinguish between a real face and a photo.
The motion based approaches~\cite{chen2014sensor,li2015seeing,ChenWZZ20} focus on analyzing the consistency of movement with the auxiliary of the inertial sensor by evaluating if 
the data collected from the inertial sensor.
Unfortunately, such defenses are very vulnerable to 3D mask based attacks~\cite{erdogmus2014spoofing}, where the 3D mask naturally has spatial features as a live person. 
For instance, Xu~\etal~\cite{XuPFM16} proposed a virtual reality (VR)-based attack which does not even require the existence of a mask, and the 3D face reconstructed from the images of the victim in VR devices 
can bypass the motion based approach~\cite{li2015seeing}. 

\noindent{\bf Texture Pattern Based Approaches.} 
The approaches distinguish genuine faces from fake faces by analyzing the texture features~\cite{tan2010face,kim2012face,boulkenafet2016face,peng2018face} since human skin, paper, and liquid-crystal display (LCS) are with different material features. 
Local Binary Pattern (LBP)~\cite{OjalaPM02} is a powerful description method which is usually utilized in these approaches to extract texture features. 
However, these defense methods are very susceptible to illumination or motion blur.
Moreover, a high-resolution screen or a realistic ``human skin'' mask may also bypass the texture verification. Recently, deep learning methods~\cite{MenottiCPSPFR15,george2019biometric,george2020learning} are widely explored for face detection, which significantly simplifies feature extraction. Unfortunately, real-word adversarial attacks~\cite{sharif2016accessorize,zhou2018invisible,Komkov2019AdvHat} can still bypass the liveness detection approaches above.

{\noindent{\bf Acoustic Signals Based Techniques.} Existing acoustic signals methods for voice liveness detection~\cite{zhang2016voicelive,zhang2017hearing,wang2019voicepop} and lip reading based authentication~\cite{lu2019lippass,tan2018silentkey} can be potentially leveraged to implement face liveness detection. 
For example, VoiceLive~\cite{zhang2016voicelive} leverages the time difference of arrival (TDoA) of the voice to the phone's two microphones, Zhang~\etal~\cite{zhang2017hearing} captures the articulatory gesture of users when speaking a passphrase by ultrasonic reflections, and VoicePop~\cite{wang2019voicepop} detects the pop noise recorded in the voice due to the exhalation. However, they are unable to detect sophisticated attacks (\eg, the adversarial attack and the 3D mask attack) against face liveness detection because they used fixed signal patterns that can be easily manipulated. Moreover, they require users to be close to the device (\ie, less than 18cm) and thus cannot be applied to face liveness detection.}

%
%

\section{A Study Of Existing Systems}\label{sec:attack}

In this section, we first present our proposed 3D projection attack, which aims to invalidate existing facial recognition services and liveness detection methods. By utilizing the attack, we evaluate the security 
of the most popular commercial facial recognition services (\eg, Microsoft~\cite{Microsoft}, Amazon~\cite{Amazon}, and Face++~\cite{Face++}), and existing liveness detection methods against it. For simplicity, we use the false accept rate as the metric to show the effects of the 3D projection attack.

\begin{figure}[!t]
\centering
\includegraphics[width=0.6\columnwidth]{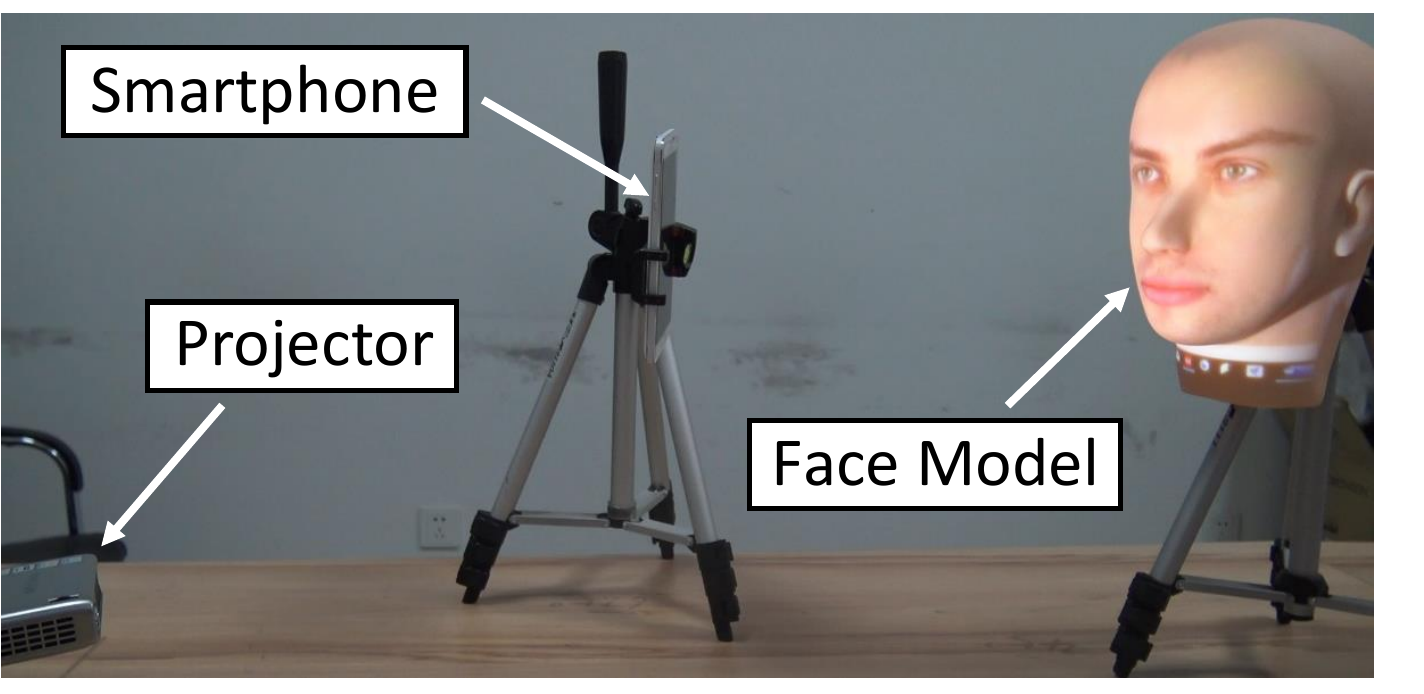}
\vspace{-2mm}
\caption{The example setup of the 3D projection attack. A reenacted facial video is projected on the silicone face model to construct the attack.}
\label{fig:projection-attacks}
\vspace{-3mm}
\end{figure}

\subsection{The 3D Projection Attack}\label{sec:projection attack}
\begin{figure}[!t]
\centering
\includegraphics[width=0.8\linewidth]{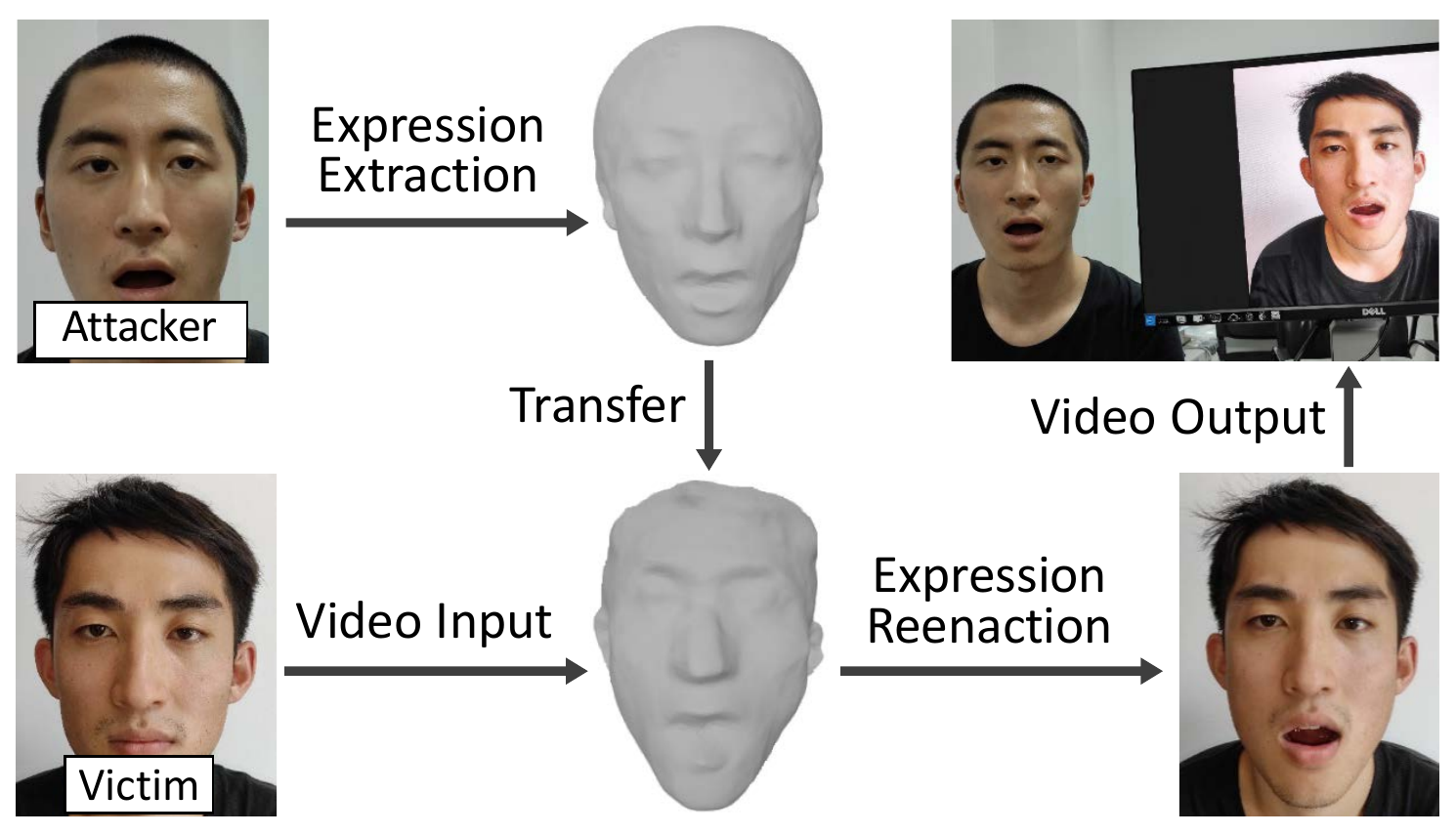}
\vspace{-2mm}
\caption{Facial mimicry through face reenactment.}
\label{fig:Against-Challenge-response}
\vspace{-3mm}
\end{figure}

\re2{Our goal is to detect a low cost but sophisticated attack that can bypass all mainstream face authentication services and typical liveness detection approaches by dynamically reconstructing the media of user face information.
We propose the 3D projection attack, which is a new type of 3D dynamic attack that is constructed with low costs and high universality.}
By only using a laptop, a mini projector, and a generic 3D silicone face model, it can bypass popular face authentication services and typical liveness detection approaches.
We conduct the attack by the following three steps.
\begin{itemize}
[leftmargin=*]
\setlength{\itemsep}{0.2pt}
  \item Step 1: We collect the photos of the victim from social networks and leverage the approach~\cite{zhu2015high} to automatically extract the landmarks of the face, with which we estimate the face pose (\ie, the angles of yaw, roll, and pitch). To obtain the front face photo of the victim, we process the photos to eliminate all deflection angles.
  \item Step 2: We precisely project the victim's face photo onto a generic 3D silicone face model. To achieve this, we adjust the distance between the mini projector and the silicone face model and keep them facing each other. We adjust the focal length to make the projected photo clear enough. The  smartphone is placed towards the face model, as shown in Fig.~\ref{fig:projection-attacks}.
  \item Step 3: We generate the corresponding responses on the basis of the challenges to pass the liveliness detection. 
  According to the study
  ~\cite{thies2016face2face}, the attacker can reenact the facial expressions of any targeted victim to create facial authentication video in real time, 
  and then the expressions will be extracted and transferred to the victim's face to synthesize the attack video, as shown in Fig.~\ref{fig:Against-Challenge-response}. Note that the attacker can also speak or move the lips. 
\end{itemize}

In our attack, the generic silicone face model and the mini projector are very cheap (i.e., only \$9 and \$35 in Amazon, respectively), which can be used for different target victims repeatedly. \red1 {The attack cost is far lower than the 3D mask attack with a custom silicone mask. Besides, we focus on the black-box attack, and the attacker does not need to have an internal perspective of the face recognition system.} Hence, the 3D projection attack is very cost-efficient, universal, and realizable.

\begin{table}[!t]
\centering
\caption{ Performances of face authentication services.}\label{tab:services-performances}
{\scriptsize
\begin{tabular}{c|| p{2cm}<{\centering} p{2.2cm}<{\centering}}
\Xhline{1.2pt}
\multirow{2}*{\footnotesize{Service system}} & \multicolumn{2}{c}{\footnotesize{Similarity (\%)}} \\
\cline{2-3}
  & \footnotesize{Legitimate} &  \footnotesize{Illegitimate} \\
\hline
\hline
\footnotesize{Microsoft face API} & \footnotesize{95.61} & \footnotesize{32.02}  \\

\footnotesize{Amazon Rekognition} & \footnotesize{99.99} & \footnotesize{30.59}  \\

\footnotesize{Face++'s face comparing} & \footnotesize{96.51} & \footnotesize{33.38} \\
\Xhline{1.2pt}
\end{tabular}
}
\end{table}

\subsection{Attacks Against Commercial Services}
Now we measure the effectiveness of the attack against commercial facial recognition services. We recruited 128 volunteers aged from 17 to 28, including 93 males and 35 females to conduct these experiments. For each volunteer, we collect three face photos from different angles. The photos can be used as 1) positive samples to evaluate the success rate of the 3D projection attack on each service, and 2) the facial source of target victims for generating 3D projection attacks videos. Then, we collect three photos of each volunteer from different angles in the 3D projection attack and use them as negative samples against each studied service.


We tested each studied service as follows. First, each volunteer uploads three of his/her genuine photos from different angles to the service for registration. The performance of each service is evaluated by measuring \qi{the similarities of legitimate attempts and illegitimate attempts, respectively. The similarity of legitimate attempts means the similarity between two arbitrary photos of the same volunteer, while the similarity of illegitimate attempts indicates the similarity between different volunteers. 
To evaluate the security of each facial recognition service against our 3D projection attack, we placed the smartphone with each service towards the projected face model of each volunteer from different angles, \ie, the left front, the front, and the right front. The results of different face authentication service systems are shown in Table~\ref{tab:services-performances}. We can see that the threshold of 50\% is effective in identifying all legitimate users and rejecting all illegitimate users since this setting can well differentiate between the similarities of legitimate attempts and that of illegitimate attempts.
Thus, we set a 50\% similarity as the threshold for correct verification, which is consistent with the existing study~\cite{uzun2018rtcaptcha}.  
}

\noindent{\textbf{Results of Attack Effectiveness}.}
\qi{
According to Table~\ref{tab:spoofing-services}, we can observe that the false accept rates of the 3D projection attack against various commercial facial recognition services are very high, which demonstrates the spoofing effects of the attack on the services.
Over 90\% spoofed faces from different angles are detected as genuine ones by facial recognition services of Face++ and Amazon. The security of Microsoft face API against the 3D projection attack is slightly stronger than that of Face++'s Face Comparing and Amazon's Recognition. However, it still achieves a 75\% success rate from the front angle.
Microsoft Face API is more sensitive to the shooting angles of photos, and the similarity of samples from large angles decreases sharply.
}


\begin{table}[!t]
\centering
\caption{Spoofing results of face authentication services.}\label{tab:spoofing-services}
\vspace{-0.5mm}
{\scriptsize
\begin{tabular}{c|| p{1.5cm}<{\centering} p{1.5cm}<{\centering} p{1.5cm}<{\centering}}
\Xhline{1.2pt}
\multirow{2}*{\footnotesize{Service system}} & \multicolumn{3}{c}{\footnotesize{False accept rate (\%)}} \\
\cline{2-4}
  & \footnotesize{Left front} & \footnotesize{Front} & \footnotesize{Right front} \\
\hline
\hline
\footnotesize{Microsoft face API} & \footnotesize{36.72} & \footnotesize{75.00}  & \footnotesize{33.59} \\

\footnotesize{Amazon Rekognition} & \footnotesize{92.97} & \footnotesize{95.31}  & \footnotesize{90.62} \\

\footnotesize{Face++'s face comparing} & \footnotesize{96.87} & \footnotesize{97.66} & \footnotesize{96.87} \\
\Xhline{1.2pt}
\end{tabular}
}
\end{table}

\subsection{Attacks Against Liveness Detection}
Now we study the effectiveness of the attack against liveness detection. The commercial services above do not enable their own liveness detection mechanisms. Therefore, we implement three types of liveness detection approaches, \ie, challenge-response based approaches, facial 3D shape based approaches, and texture pattern based approaches, and evaluate the impact of the 3D projection attack on them. 
As a comparison, we also measure the impact of the MFF attack. In our experiments, the collected face photos are used to generate attack videos to conduct the 3D projection attack and the MFF attack. The experiment setup of the 3D projection attack has been described in Section~\ref{sec:projection attack}. In MFF attacks, we record the victim's facial videos and display them on the screen.

\begin{table}[!t]
\centering
\caption{Attack results against liveness detection.}\label{tab:attack-results}
{\scriptsize
\begin{tabular}{c|| p{2.6cm}<{\centering} p{1.5cm}<{\centering}}
\Xhline{1.2pt}
\multirow{2}*{\footnotesize{Liveness detection}} & \multicolumn{2}{c}{\footnotesize{False accept rate (\%)}} \\
\cline{2-3}
  & \footnotesize{3D projection attack} &  \footnotesize{MFF attack} \\
\hline
\hline
\footnotesize{Challenge-response} & \footnotesize{97.86} & \footnotesize{1.28}  \\

\footnotesize{Facial 3D shape} & \footnotesize{92.31} & \footnotesize{0}  \\

\footnotesize{Texture pattern} & \footnotesize{30.34} & \footnotesize{2.99} \\
\Xhline{1.2pt}
\end{tabular}
}
\end{table}


\noindent{\textbf{Challenge-Response Based Approaches}.} Microsoft Face API integrates emotion recognition, which can be utilized to implement the challenge-response based liveness detection mechanism (\eg, smile detection). Therefore, we capture three photos of each generated attack video from different angles and distances and test the probability of each captured photo being understood as a smiling face by Microsoft Face API. 
A projective photo of reenacted the smile face on the generic silicone face model easily reaches a 99.9\% smiling probability by Microsoft Face API. However, a displayed photo of the original face on the screen can only get a 0.1\% smiling probability. Based on  Table~\ref{tab:attack-results}, we can see that the false accept rate of the 3D projection attack is as high as 97.86\%, while that of MFF attack is only 1.28\%. We can conclude that the 3D projection attack can defeat challenge-response based liveness detection approaches because we can generate the responses according to the challenges in real time.

\begin{figure}[!t]
\centering
\includegraphics[width=0.8\columnwidth]{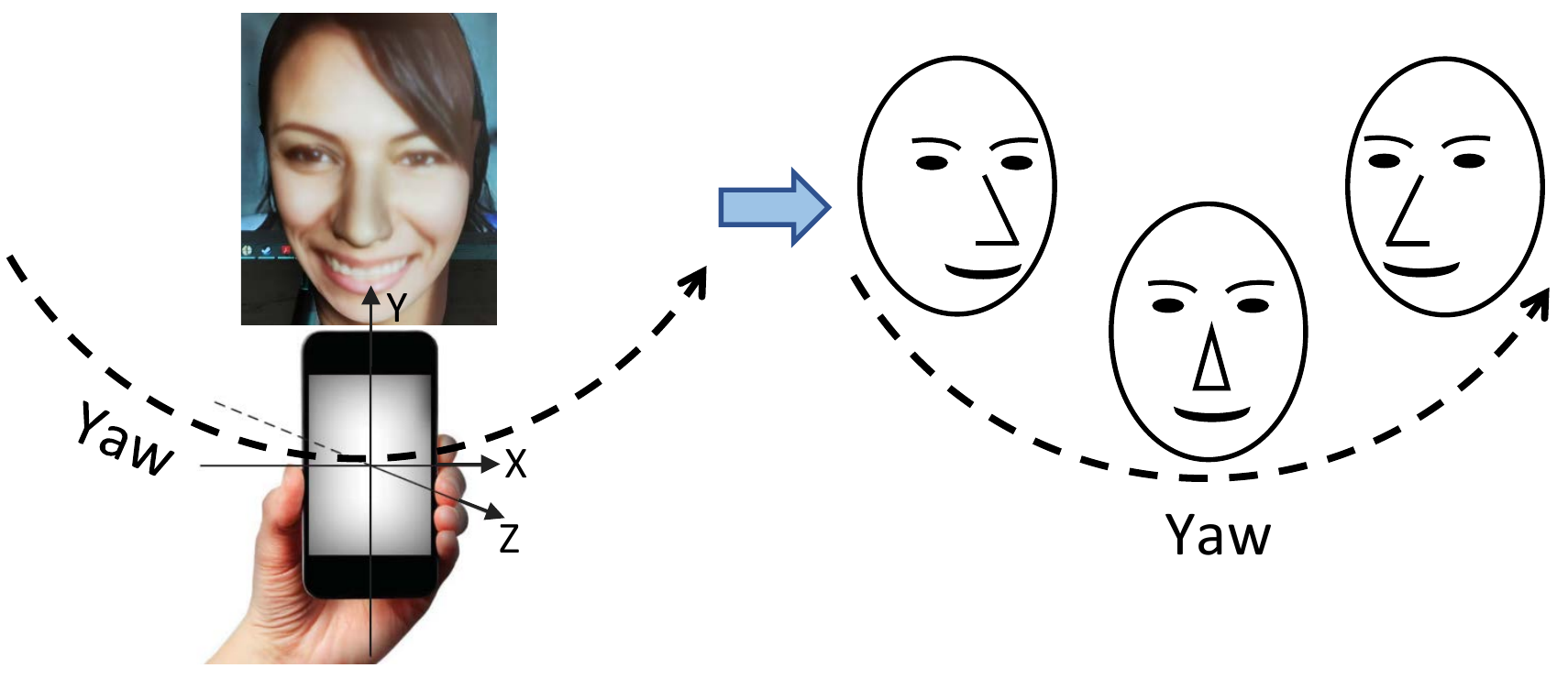}
\vspace{-2mm}
\caption{The effect of 3D projection attack on the consistency of movement check method.}
\label{fig:Against-3D-shape}
\vspace{-3mm}
\end{figure}

\noindent{\textbf{Facial 3D shape Based Approaches}.}
\qi{Similar to~\cite{chen2014sensor,li2015seeing}, we leverage the consistency of the movement check method to implement the facial 3D shape based liveness detection mechanism, which only requires a common front camera and motion sensors. Note that, it is difficult to develop 3D shape based defenses that build precise 3D models of users' faces on the mainstream smartphones since 
they usually require various expensive hardware. 
For instance, Apple's Face ID~\cite{FaceID} can only be enabled on the device equipped with an ambient light sensor, a distance sensor, an infrared lens, a flood camera, and a dot matrix projector.} As shown in Fig.~\ref{fig:Against-3D-shape}, we move the smartphone in front of the face image projected on the generic silicone model over a short distance and verify the consistency between the device movement data extracted from the motion sensors and the real changes of the nose captured by the front-facing camera. Since the victim's facial video is projected on the 3D silicone face model in our 3D projection attack, the facial structures recovered from the video contain 3D shape information. Correspondingly, in the MFF attack, the victim's facial video is displayed on the plane screen. Therefore, the recovered facial structures do not contain any 3D shape information. As a result, the false accept rate of 3D projection attack is as high as 92.31\%, while that of MFF attack is 0. We can conclude that the 3D projection attack can also defeat facial 3D shape based liveness detection approaches  due to its facial 3D structure.

\begin{figure*}[!t]
\begin{minipage}[t]{0.65\linewidth}
\centering
\includegraphics[width=0.99\textwidth]{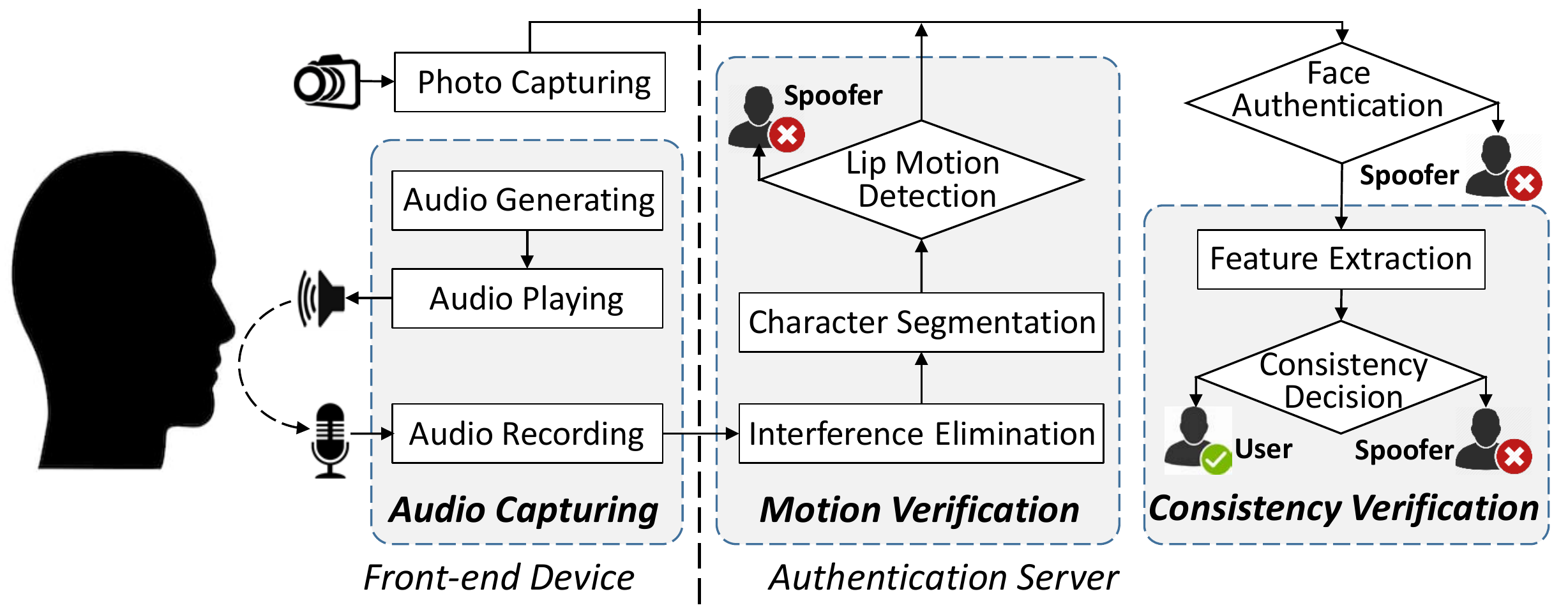}
\vspace{-2mm}
\caption{\red1{The workflow of FaceLip.}}
\label{fig:FaceLip-overview}
\end{minipage}
\begin{minipage}[t]{0.34\textwidth}
\centering
\includegraphics[width=0.99\columnwidth]{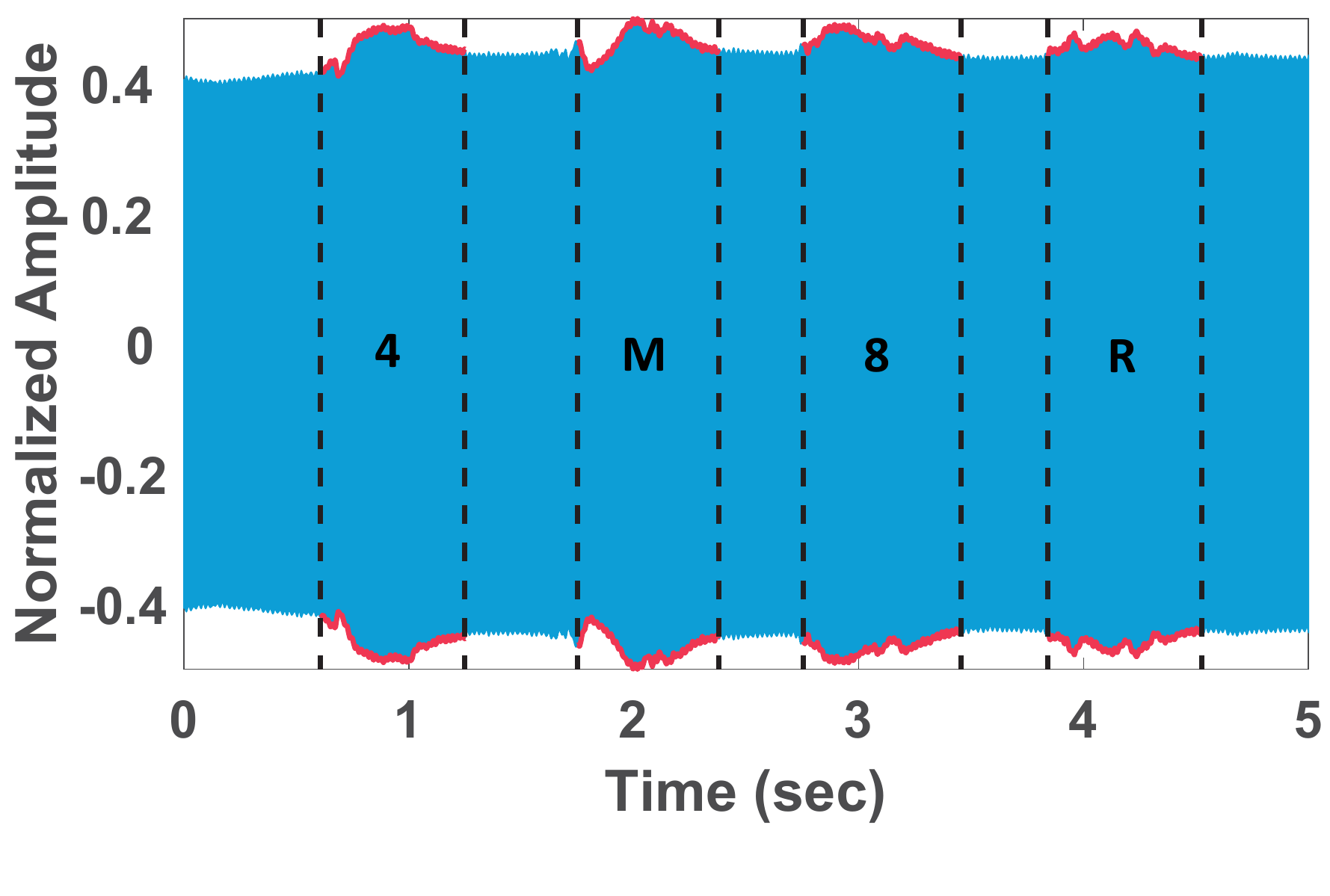}
\vspace{-6mm}
\caption{Recorded signals when reading ``4M8R''.}
\label{fig:recorded-audio}
\end{minipage}
\vspace{-3mm}
\end{figure*}

\noindent{\textbf{Texture Pattern Based Approaches}.} As mentioned above, LBP~\cite{OjalaPM02} is a powerful description method which is widely used to extract texture feature for liveness detection. Thus, we first capture three photos of each generated attack video from different angles and distances, then utilize LBP for feature extraction and Support Vector Machine (SVM) for classification. As shown in Table~\ref{tab:attack-results}, the false accept rate of MFF attack is only 2.99\%, while that of our 3D projection attack can achieve 30.34\%. The LBP method analyzes color and micro textures extracted from the captured face photos, whose detection accuracy is affected by the quality of the facial media,~\eg, the resolution, motion blurring, and shading. We can infer that it is hard to defend the 3D projection attack by using texture pattern based approaches with the high-resolution projector and realistic silicone face model, because the texture and reflectance details are similar to that of human skin.

\noindent \textbf{Summary.} \qi{In this paper, we utilize the proposed 3D projection attack to evaluate the security of popular face authentication services, \eg, Microsoft, Amazon, and Face++, and typical liveness detection approaches. 
We find that the new 3D projection attack can easily evade these face authentication services and liveness detection approaches. In order to enhance the security of face authentication systems, in the next section, we will propose a novel face liveness detection scheme, \ie, FaceLip, which is robust and secure under wide-ranging sophisticated attacks (including the 3D projection attack).
}


\section{FaceLip Design}\label{sec:lipdefence}

\subsection{An Overview}
\qi{In order to ensure strong security of face liveness detection, FaceLip utilizes the unique lip motion of each individual~\cite{faraj2006motion} to construct an unforgeable pattern that is in the form of acoustic signals generated  according to random challenges in each round of detection. We transform a phone's acoustic system into an active sonar and then compute fine-grained estimation of subtle lip motions to capture the unforgeable patterns of lip motions. In particular, we leverage the phase shift based approach, which allows us to process the reflected acoustic signals as phase modulated signals to measure the corresponding lip motions.}
Fig.~\ref{fig:FaceLip-overview} presents the workflow of FaceLip, which mainly consists of three modules: \red1{\textit{Audio Capturing}}, \textit{Motion Verification}, and \textit{Consistency Verification}. \red1{\textit{Audio Capturing}} is performed by the front-end device (\ie, smartphone), while \textit{Motion Verification} and \textit{Consistency Verification} composing our liveness detection scheme are set up in the server. \red1{In addition, \textit{Photo Capturing} and \textit{Face Authentication} are deployed in the front-end device and the server, respectively. We first perform the \textit{Motion Verification} to detect whether the recorded signals contain ``physical'' lip motions and then perform the \textit{Consistency Verification} to analyze  the entire lip motion patterns for consistency decision. Only if a user passes all these modules, FaceLip considers him/her as the real user. \re2{Note that, according to face authentication scenarios, \textit{Consistency Verification} can be performed before or after the face authentication process. If \textit{Consistency Verification} is not informed with the user identity, face authentication needs to be performed to obtain the user identity before comparing the consistency of lip motions.} 
FaceLip is orthogonal to face authentication and can be integrated into all existing face authentication systems, 
and thus we will not describe the  functionalities performed in face authentication, \eg, photo capturing and face authentication, in this paper.}

\noindent{\textbf{Audio Capturing}.}
\red1{This module transforms a phone's acoustic system into an active sonar and captures a user's unforgeable lip motion pattern via imperceptible acoustic signals. The audio signals are produced by the generator, which generates the audio signals with random carrier frequencies for liveness detection.
During the detection, we use 
the phone's speaker to play imperceptible audio and the microphone to record the acoustic signals reflected by the speaking lips at the same time. 
Then, the recorded acoustic signals are uploaded to the server for liveness detection.
}

\noindent{\textbf{Motion Verification}.}
\reda3{This module detects whether the recorded acoustical signals contain valid information of lip motions, which can defeat the typical MFF attacks}. To achieve the goal, this module includes three steps: 
\textit{Interference Elimination} removes the noise in the recorded acoustical signals to obtain pure signals reflected by the lips, which allows us to accurately capture users' lip motions even at larger distances (\eg, larger than 18cm). It solves the key challenge of using acoustic signals for face liveness detection on the smartphone.
\textit{Character Segmentation} segments the acoustic signals into fragments corresponding to each character, and \textit{Lip Motion Detection} detects whether the fragments contain ``physical'' lip motions. \red1{The information of lip motions extracted from the acoustic signals can only be produced from ``physical'' movements rather than ``visual'' movements, while the MFF attacks only contain ``visual'' lip motions. Therefore, this module can effectively resist MFF attacks.}


\noindent{\textbf{Consistency Verification}.}
\red1{This module verifies the consistency of lip movements for detection and that collected during the user register phase, which can defeat sophisticated attacks, \ie, the 3D dynamic attacks. To achieve this, FaceLip extracts robust energy-band time-frequency features from the acoustic fragments containing the information of lip motions. Then, we leverage the lip motion features extracted during user registration to construct the user profile and build a binary classifier for the user, which is used to decide whether each input of liveness detection is consistent with the real user.
Due to the uniqueness of the lip motion patterns of different individuals~\cite{faraj2006motion}, the \textit{Consistency Verification} module can effectively resist 3D dynamic attacks.}

\subsection{Audio Capturing}

FaceLip first generates imperceptible audio consisting of several tones to capture unforgeable lip motion patterns, whose carrier frequencies are controlled by the randomizer. After the audio is produced by the generator, FaceLip starts to use the speaker of the device to play the imperceptible audio and the microphone to record the acoustic signals reflected by the lips.

\noindent{\textbf{Audio Generating}.} The generated audio is the superposition of multiple tones $\sum \nolimits_{i=1}^N 2A\cos 2\pi f_i t$, where $2A$ is the amplitude, $f_i$ is the carrier frequency of acoustic signals, and $N$ is the total number of carrier frequencies. In our design, $N$ is set as 3, which results in 3 tones with different frequencies. We leverage the randomizer to generate the frequency $f_i$. To avoid the interference from adjacent frequencies, we set the frequency interval of any two tones $\Delta f$ to be at least 300Hz. We randomize the carrier frequencies of the generated audio to resist the audio replay attack. Therefore, the attacker cannot bypass our liveness detection by replaying pre-recorded audio that was collected during previous rounds of detection. \qi{Note that, most people cannot hear the tone whose frequency is higher than 18KHz~\cite{zhou2019stealing} and the performance of acoustic components in some smartphones decreases significantly when the frequency is higher than 21KHz~\cite{zhou2018enabling}. Thus, we set $f_i$ in the range of $18 \sim 21$KHz.}

\noindent{\textbf{Audio Playing and Recording}.} 
\red1{The user is required to speak the passcode during liveness detection\footnote{The user can speak aloud or silently. However, for the security of the passcode, we suggest speaking silently.}. \re2{The passcode can consist of different characters, including numbers (\ie, $0\sim9$) and letters with obvious lip motions (\eg, F, H, \etc).
The passcode is initially set up during the user registration phase and can be changed by users. 
FaceLip will judge the applicability of each character in each user's passcode to ensure that the character  involves more obvious lip motions.
}
In the meantime,} FaceLip uses the phone's speaker to play the generated imperceptible audio and the microphone to record the acoustic signals. The recorded signals contain the information of lip motions because the speaking lips reflect the played audio. Fig.~\ref{fig:recorded-audio} shows an example of recorded acoustic signals when the user speaks the passcode. We can observe that the lip motions will lead to unique patterns on the acoustic signals (\ie, signals in the red wireframes shown in Fig.~\ref{fig:recorded-audio}).
\red1{Then, the audio is transmitted from the front-end smartphone to the back-end server for the following processing. The communication between the smartphone and the server is protected by standard network protocols, \eg, HTTPs. }

\begin{figure}[!t]
\centering
\includegraphics[width=0.8\columnwidth]{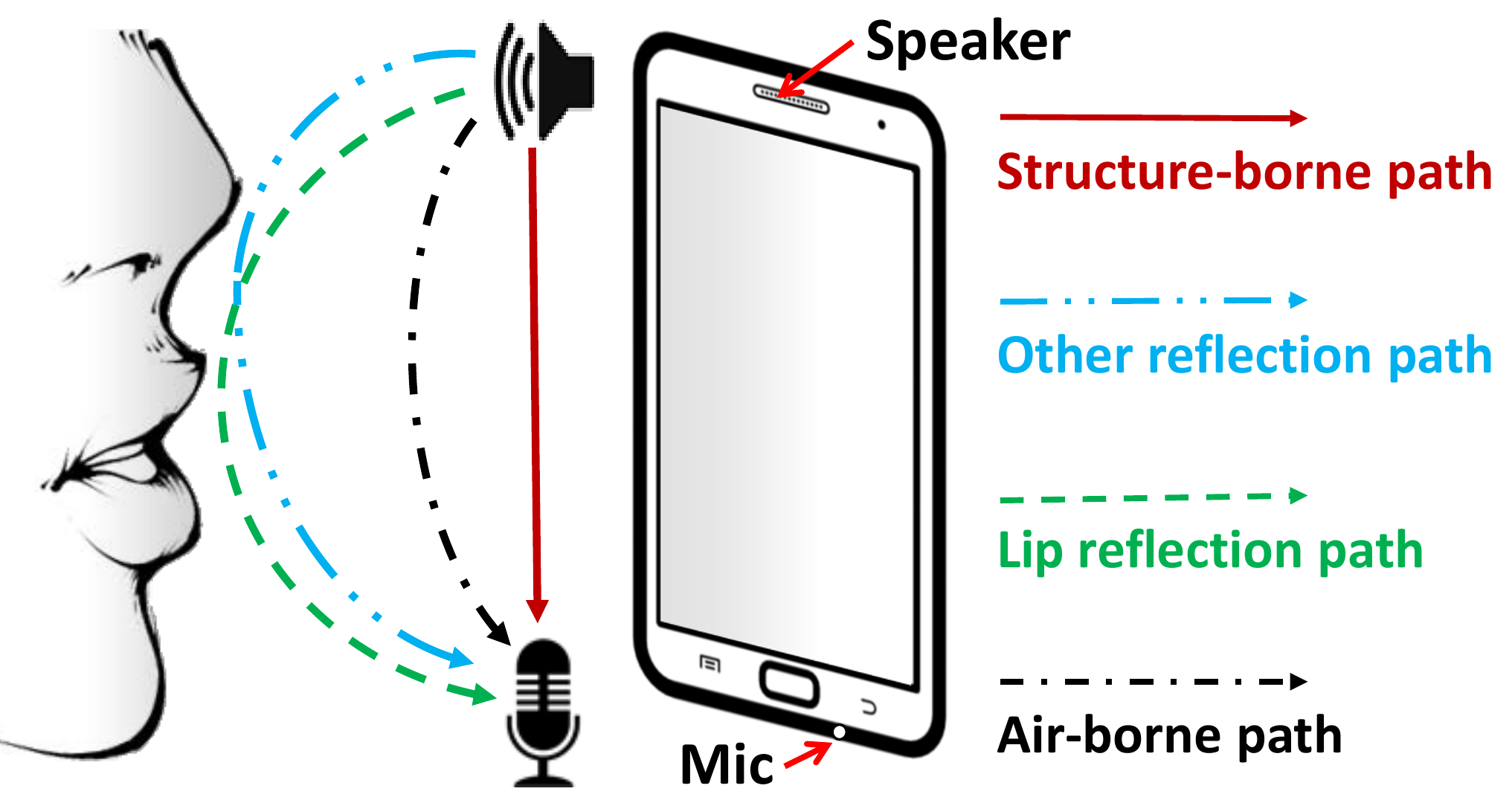}
\vspace{-2mm}
\caption{Multiple propagation paths for the acoustic signal.}
\label{fig:different-paths}
\vspace{-3mm}
\end{figure}

\begin{figure*}[!t]
\begin{minipage}[t]{0.327\linewidth}
\centering
\subfigure{
\includegraphics[width=0.9\textwidth]{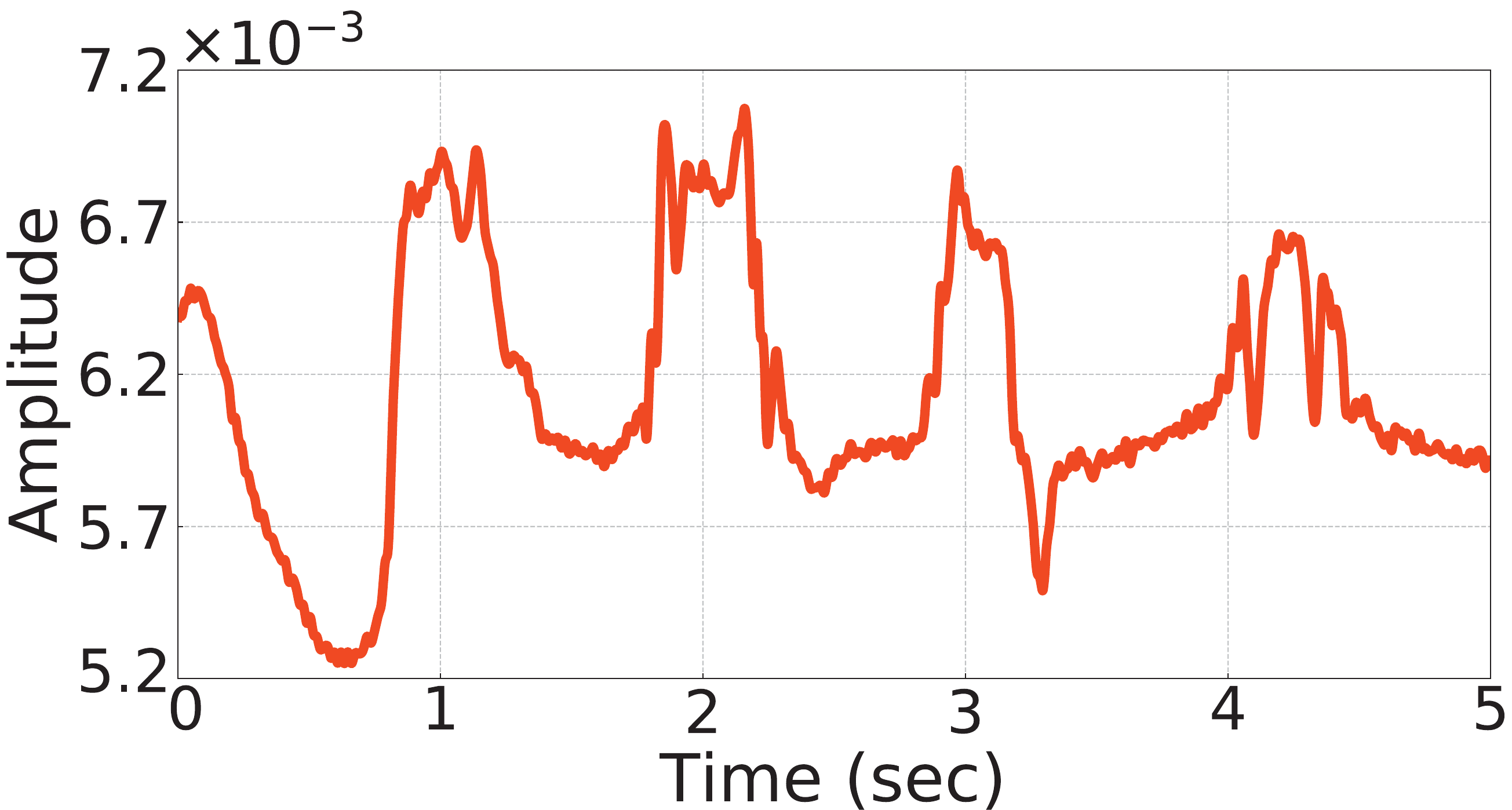}}
\end{minipage}
\begin{minipage}[t]{0.327\linewidth}
\centering
\subfigure{
\includegraphics[width=0.95\textwidth]{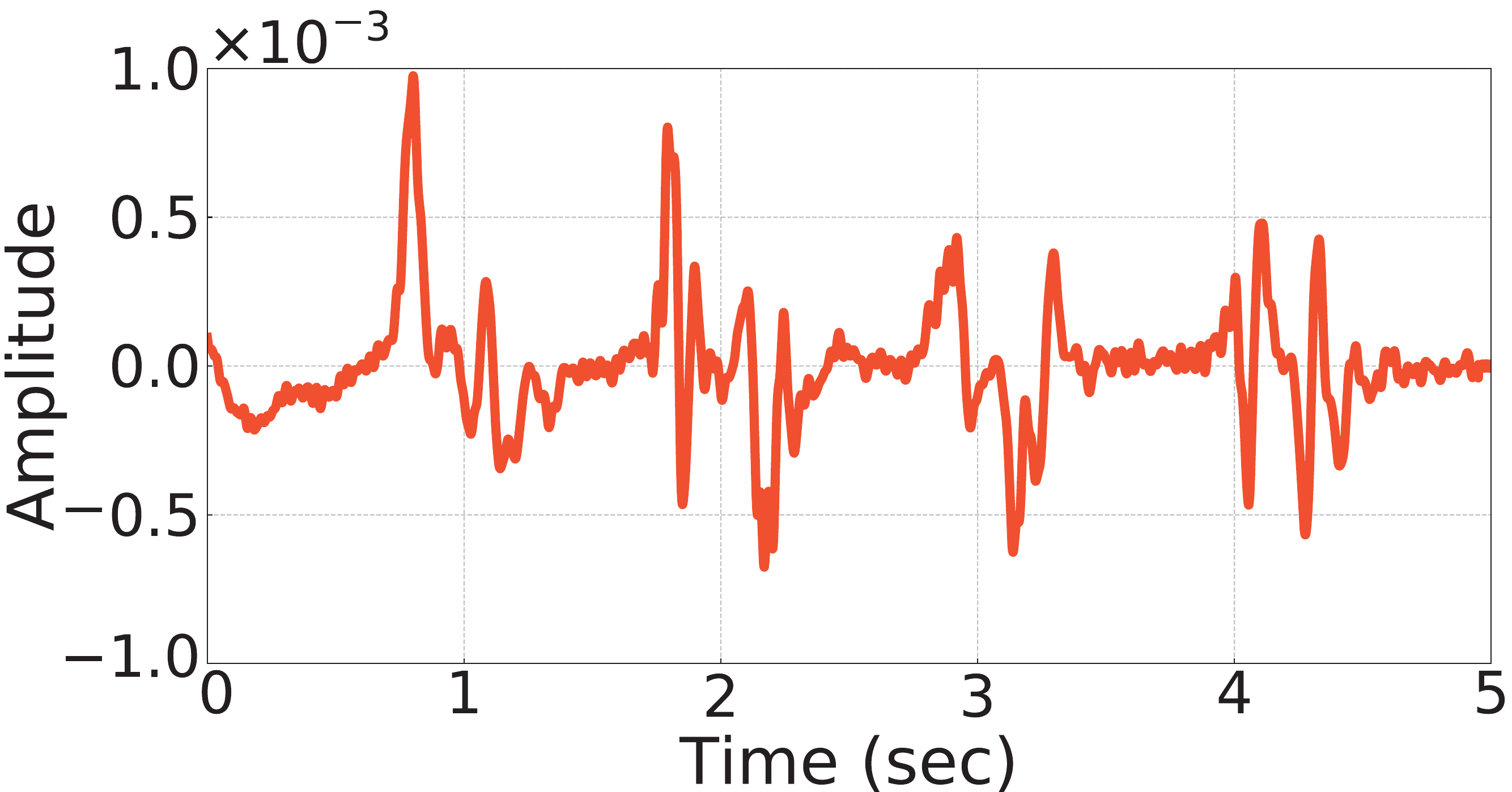}}
\end{minipage}
\begin{minipage}[t]{0.327\linewidth}
\centering
\subfigure{
\includegraphics[width=0.95\textwidth]{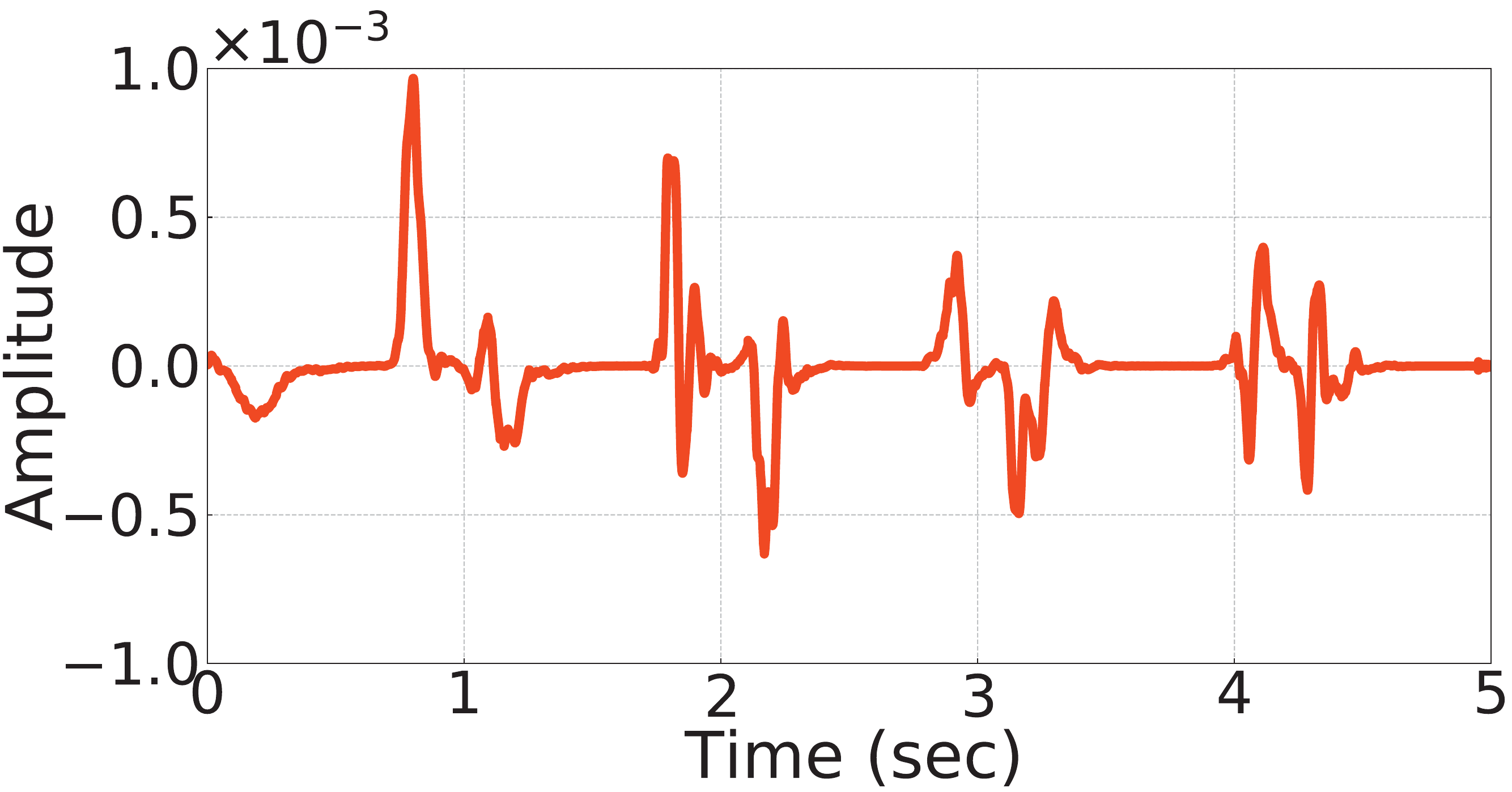}}
\end{minipage}
\vspace{-5mm}
\end{figure*}
\begin{figure*}[!t]
\begin{minipage}[t]{0.327\linewidth}
\centering
\subfigure{
\includegraphics[width=0.9\textwidth]{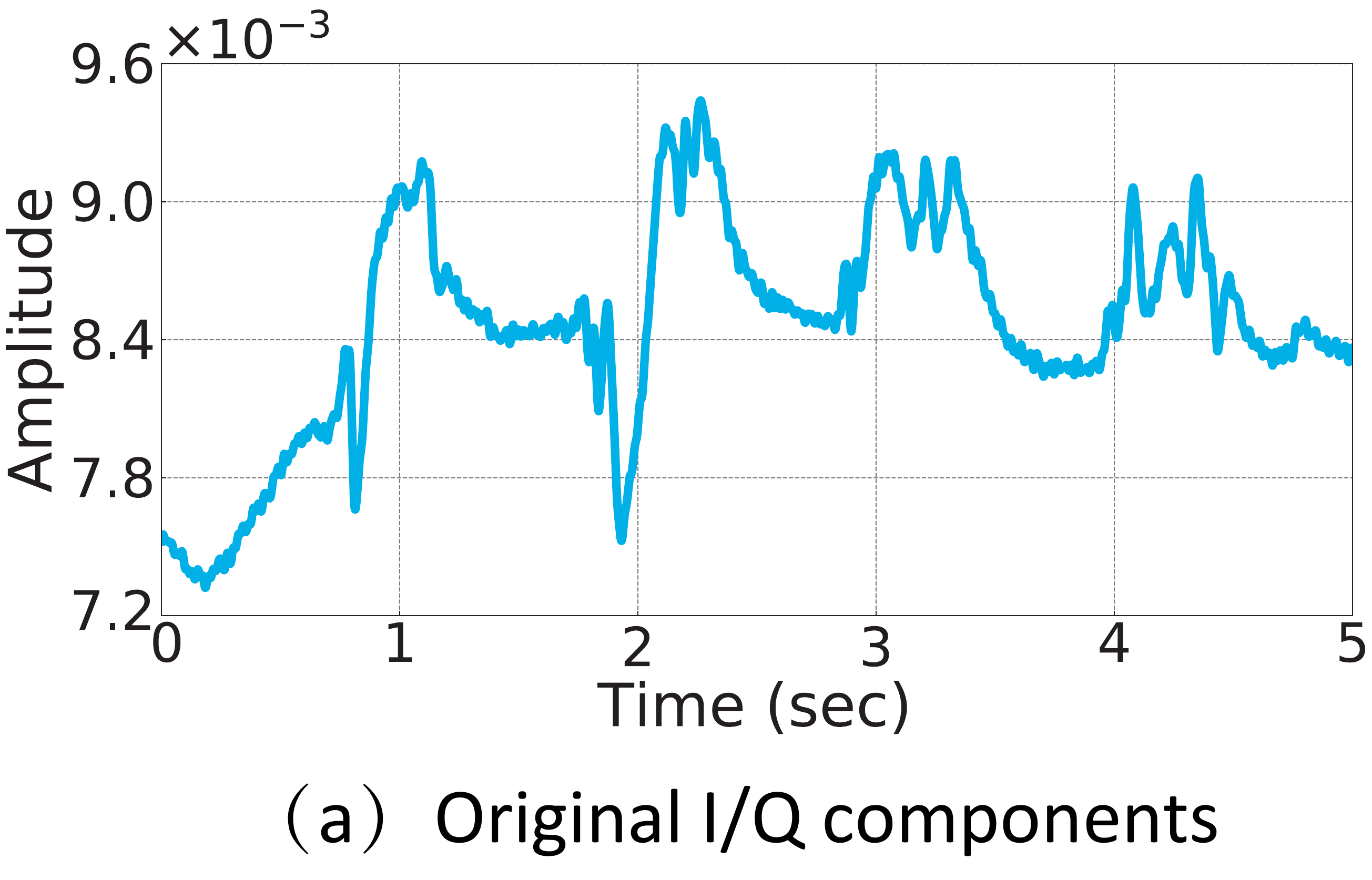}}
\end{minipage}
\begin{minipage}[t]{0.327\linewidth}
\centering
\subfigure{
\includegraphics[width=0.95\textwidth]{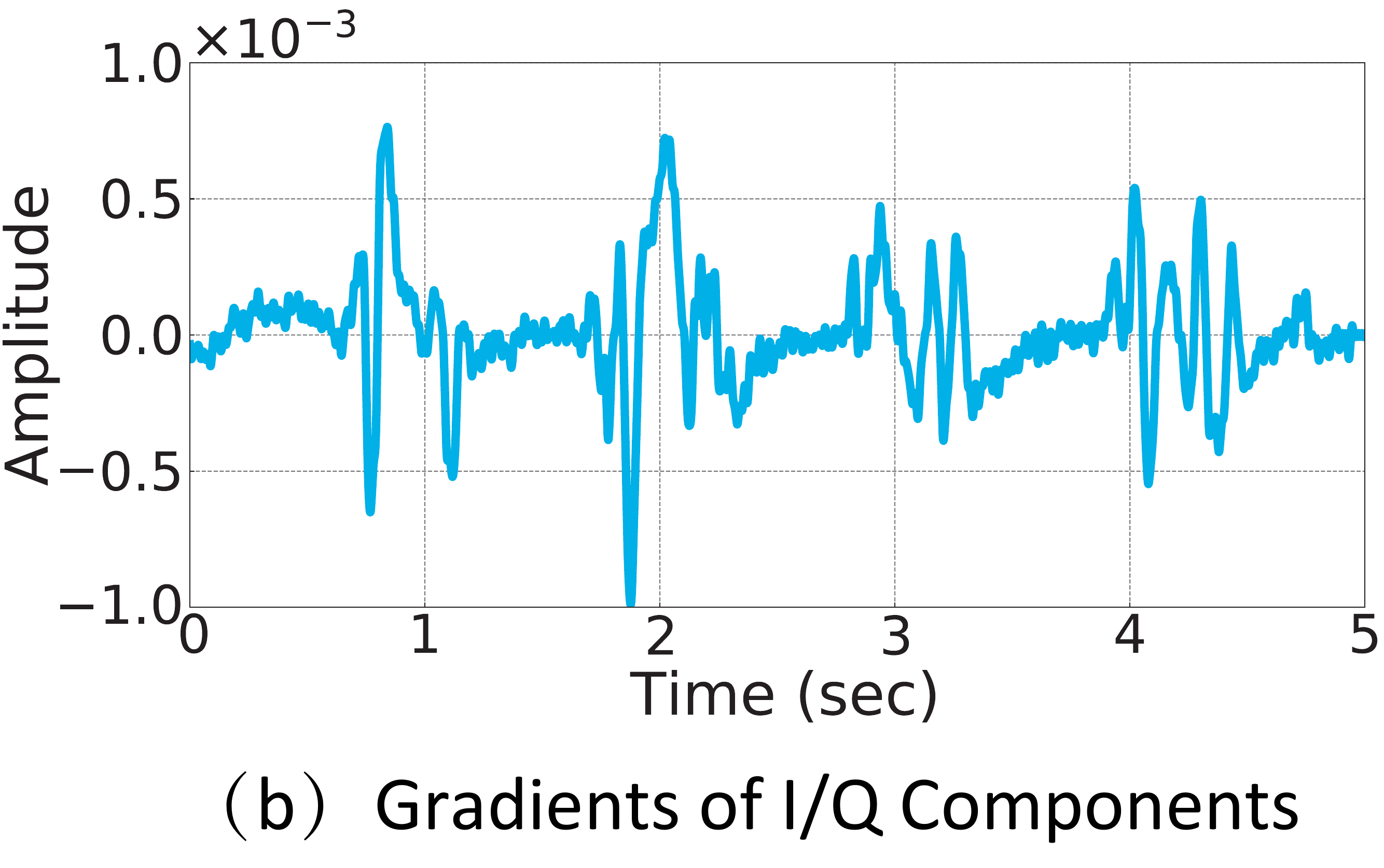}}
\end{minipage}
\begin{minipage}[t]{0.327\linewidth}
\centering
\subfigure{
\includegraphics[width=0.95\textwidth]{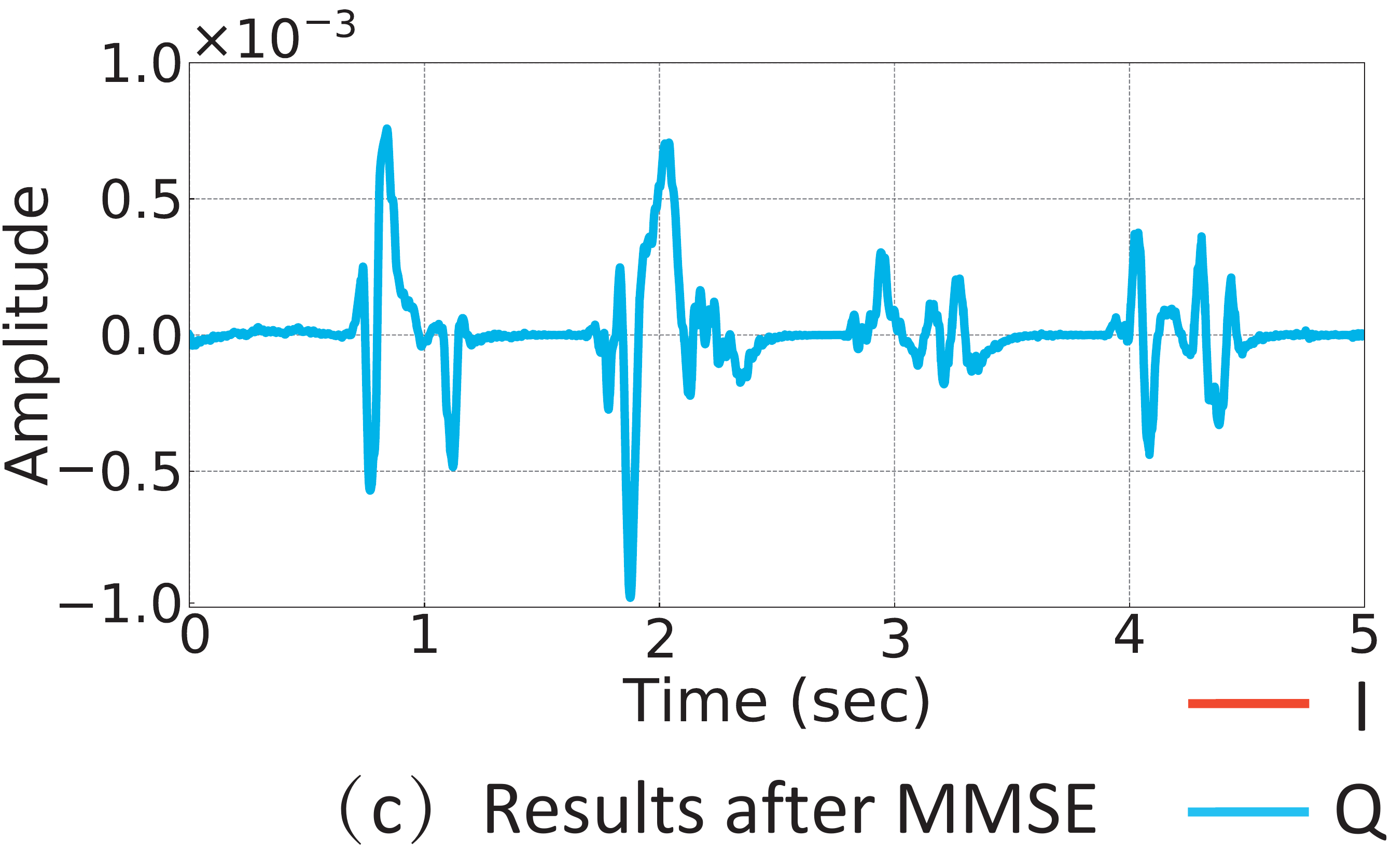}}
\end{minipage}
\vspace{-2mm}
\caption{The results of each step in static interferences elimination.}
\label{fig:interference-removal}
\vspace{-3mm}
\end{figure*}


\subsection{Motion Verification}
Motion verification aims to detect if the detected motions are generated by real users. In order to efficiently and correctly analyze the signals, we eliminate multi-path interference in the signals and then segment the signals corresponding to each character in the passcode.

\noindent \textbf{Interference Elimination.} In this step, FaceLip first leverages the coherent detection 
to down-convert the recorded audio to demodulate the baseband signals and then eliminates the multi-path interference to obtain the acoustic signal components only related to lip motions.

\noindent{\textit{Coherent Detection}.}
As shown in Fig.~\ref{fig:different-paths}, there are multiple propagation paths from the speaker to the microphone for the audio during face liveness detection, including the reflection path from the user's lips, the structure-borne path, the air-borne path, and the reflection paths from surrounding objects (\eg, the user's face). Let $k$ denote the $k$-{th} path, $2A_{k}(t)$ denote the amplitude of acoustic signals in the $k$-{th} path, $2\pi f_i \frac{d_{k}(t)}{\nu_k}$ denote the phase shift introduced by propagation delays, and $\theta_{k}$ be the phase shift caused by system delays. Assuming that there are $M$ paths, the recorded audio $R(t)$ can be described as
    \begin{equation}\label{received signal}
        \begin{aligned}
        R(t) = \sum\limits_{k=1}^M \sum\limits_{i=1}^N 2A_{k}(t)\cos{(2\pi f_i t-2\pi f_i \frac{d_{k}(t)}{\nu_k}-\theta_{k})}.
        \end{aligned}
    \end{equation}

FaceLip regards the original audio played by the speaker as the carrier signals and treats the recorded audio by the microphone $R(t)$ as the superposition of several baseband signals modulated with the phase shift. As mentioned above, the generated audio is the superposition of multiple tones with different frequencies. The tones with different frequencies can be regarded as different baseband signals. We first leverage the band-pass filter with the frequency band $[f_i-\Delta f/2, f_i+\Delta f/2]$ to segment the received signals into $N$ baseband signals. The $i$-{th} baseband signals can be denoted as $R_i(t) = \sum\nolimits_{k=1}^M 2A_{k}(t)\cos{(2\pi f_i t-2\pi f_i \frac{d_{k}(t)}{\nu_k}-\theta_{k})}$.

The phase shift of $R_i(t)$ remains constant for the structure-borne path, the air-borne path, and the reflection paths from static objects, while the phase shift for the reflection path from the speaking lips and surrounding dynamic objects is varied due to the changes of propagation delay.
Since the recorded signals are synchronized with the played signals, FaceLip can leverage the coherent detector to demodulate the recorded acoustical signals to obtain the I (In-phase) component and Q (Quadrature) component of the baseband signals. Let $F_{l}$ denote a low pass filter, and $F_{d}$ denote a down-sampled function. The corresponding I and Q components demodulated by the coherent detection can be represented as
    \begin{equation}\label{eq:I/Q}
        \begin{aligned}
        I(t) &= \sum\limits_{i=1}^N (F_{d}(F_{l}(R_i(t) \times  \cos 2\pi f_i t))), \\
        Q(t) &= \sum\limits_{i=1}^N (F_{d}(F_{l}(R_i(t) \times (-\sin 2\pi f_i t)))),
        \end{aligned}
    \end{equation}
where $R(t)$ is multiplied with the original acoustic signals $\cos 2\pi f_i t$. We have
    \begin{equation}\label{coherent detector}
        \begin{split}
            R_i(t) \times \cos 2\pi f_i t &=  \sum\limits_{k=1}^M  A_{k}(t) \{\cos{(2\pi f_i \frac{d_{k}(t)}{\nu_k}+\theta_{k})} \\
           &+\cos{(4\pi f_i t - 2\pi f_i \frac{d_{k}(t)}{\nu_k}-\theta_{k})}\}.
        \end{split}
    \end{equation}
The product is then filtered by the low pass filter $F_{l}$ to remove the $2f_i$ high-frequency term in Eq.~\eqref{coherent detector} and down-sampled by $F_{d}$.
Then the I component of the baseband signals is calculated as $I(t) = \sum\nolimits_{k=1}^M \sum\nolimits_{i=1}^N A_{k}(t) \cos{(2\pi f_i \frac{d_{k}(t)}{\nu_k}+\theta_{k})}$.
Similarly, the Q component can be calculated as $Q(t) = -\sum\nolimits_{k=1}^M \sum\nolimits_{i=1}^N A_{k}(t) \sin{(2\pi f_i \frac{d_{k}(t)}{\nu_k}+\theta_{k})}$.

\noindent{\textit{Dynamic Interference Elimination}.}
After coherent detection, the demodulated baseband signals $I(t)$ and $Q(t)$ are still the superposition of multi-path acoustic signals. To obtain the signals only related to the lips, it is necessary to remove the interference signals from other paths. The multi-path interference includes the dynamic interference with varied phase shift and the static interference with constant phase shift. Dynamic interference involves the signals reflected by other nearby moving objects, \eg, the user's hand. Static interference involves signals from the structure-borne path and the air-borne path, as well as the signals reflected by nearby static objects.

In most scenarios, nearby moving objects are the human bodies, which usually cause a frequency shift in the
range of $50\sim200$Hz~\cite{SBM2012}. While the maximum frequency shift induced by lip motions usually does not exceed 40Hz~\cite{tan2017silenttalk,WuYZCW19}.
Therefore, FaceLip can leverage the low-pass filter to eliminate the dynamic interference. The cut-off frequency of the low-pass filter $F_{l}$ in \textit{Coherent Detection} is set as 40Hz to eliminate the dynamic interference and demodulate baseband signals simultaneously.

\noindent{\textit{Static Interference Elimination}.} After dynamic interference elimination, the acoustic signals are the superposition of signals reflected by the lips and static interference signals. To remove the interference, we regard I/Q components as the sum of static component $I_{s}(t)/Q_{s}(t)$ which is a constant and acoustic signal reflected by the speaking lips as
\begin{equation}\label{static interferences removal}
    \begin{aligned}
       I(t) &= I_{s}(t) + \sum\limits_{i=1}^N A_{l}(t) \cos{(2\pi f_i \frac{d_{l}(t)}{\nu}+\theta_{l})},\\
       Q(t) &= Q_{s}(t) - \sum\limits_{i=1}^N A_{l}(t) \sin{(2\pi f_i \frac{d_{l}(t)}{\nu}+\theta_{l})},
    \end{aligned}
\end{equation}
where $A_{l}(t)$ is the amplitude of signals reflected by the lips, $d_{l}(t)$ is the propagation delay, $\nu$ is the propagation velocity of sound in the air, and $\theta_{l}$ is the phase shift caused by the system delay. For simplicity, we use $\phi_{l}(t)$ to indicate $2\pi f_i \frac{d_{l}(t)}{\nu} +\theta_{l}$. Thus, the I component of superposed baseband signals can be denoted as $I(t) = I_{s}(t) + \sum\nolimits_{i=1}^N A_{l}(t) \cos{(\phi_{l}(t))}$, and the Q component of superposed baseband signals can be denoted as $Q(t) = Q_{s}(t) - \sum\nolimits_{i=1}^N A_{l}(t) \sin{(\phi_{l}(t))}$.

To eliminate the static component, we calculate the gradient of I/Q component $I_{g}(t)/Q_{g}(t)$ as
\begin{equation}\label{gradient}
    \begin{aligned}
       I_{g}(t) &= A'_{l}(t) \cos{(\phi_{l}(t))} - A_{l}(t)\phi'_{l}(t)\sin{(\phi_{l}(t))},\\
       Q_{g}(t) &= -A'_{l}(t) \sin{(\phi_{l}(t))} - A_{l}(t)\phi'_{l}(t)\cos{(\phi_{l}(t))},
    \end{aligned}
\end{equation}
where $A'_{l}(t)$ and $\phi'_{l}(t)$ are the derivative of $A_{l}(t)$ and $\phi_{l}(t)$ respectively. $A_{l}(t)$ is a coefficient inversely proportional to the square of the propagation distance.
Since user's lip motions are very subtle, the value of $A_{l}(t)$ hardly varies with the lip movement, which means that the value of $A'_{f}(t)$ approximates zero.
Thereby, $I_{g}(t)$ can be denoted as $- A_{l}(t)\phi'_{l}(t)\sin{(\phi_{l}(t))}$, and $Q_{g}(t)$ can be denoted as $- A_{l}(t)\phi'_{l}(t)\cos{(\phi_{l}(t))}$.


Considering its outstanding performance in dealing with the random error, we further use the \emph{minimum mean square error} (MMSE)~\cite{MMSE} to eliminate the slowly-changing term of $I_{g}(t)$ and $Q_{g}(t)$. Fig.~\ref{fig:interference-removal} shows the results of each step in \textit{Static Interference Elimination}. It can be seen that the magnitudes of $I_{g}(t)$ and $Q_{g}(t)$ are close to zero in the absence of lip movements after MMSE.

\begin{figure}[!t]
\centering
\includegraphics[width=0.7\linewidth]{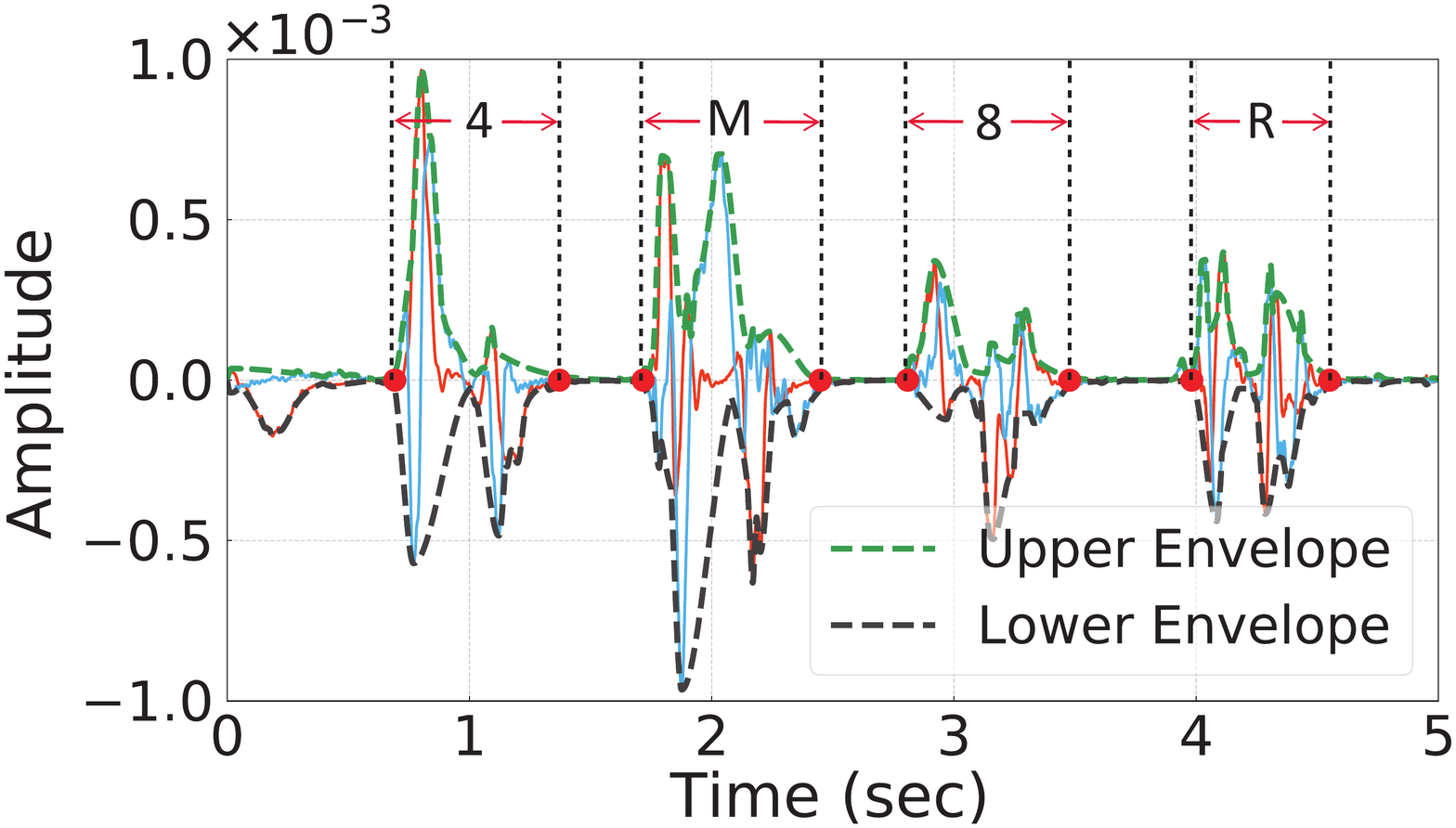}
\vspace{-2mm}
\caption{An example of character segmentation.}
\label{fig:line-seg}
\vspace{-3mm}
\end{figure}

\noindent \textbf{Character Segmentation.}
To detect valid lip motions, we need to segment signals into fragments corresponding to each character of the passcode. In FaceLip, we develop a two-step segmentation algorithm to realize character segmentation. Specifically, we first leverage the Voice Activity Detection (VAD) algorithm provided by the VOICEBOX~\cite{VOICEBOX} toolbox to roughly locate the lip motion fragments and then segment characters based on the signal envelope.

During liveness detection, the user is required to speak the passcode consisting of several characters. The key issue of character segmentation is how to locate the start and end signal points of each character. Fig.~\ref{fig:interference-removal}(c) shows that there exists a short silent time when the lips are relatively motionless between two adjacent characters. As a result, the I/Q waveform of signals fluctuates slowly during the silent time. 
Based on this observation, we propose a VAD together with the envelope-based lip motion segmentation algorithm to accurately identify the start and end signal points of each character.

FaceLip first performs the VAD algorithm on $I_{g}(t)$ and $Q_{g}(t)$ to roughly locate the fragments, which include the start and end signal points of each character. To  locate them accurately, FaceLip then estimates the upper and lower envelopes of $I_{g}(t)$ and $Q_{g}(t)$. As shown in Fig.~\ref{fig:line-seg}, the difference between upper and lower envelopes changes with time. The value of the difference becomes larger when the lips open and approaches zero when the lips close. Therefore, FaceLip sets a threshold $T_d$ for the difference and a threshold $T_{w}$ for the duration of a character and goes across the difference over time. When the difference becomes larger than $T_d$, the signal point may be the start signal point of a character. Correspondingly, when the difference becomes smaller than $T_d$, and the interval between the current and the start signal points is longer than $T_{w}$, this signal point may be the end signal point of a character. Through repeated testing, we adjust the value of $T_d$ and $T_{w}$ to confirm the start and the end signal points of each character accurately. \re2{Note that character segmentation is a necessary and lightweight step to preprocess the acoustic signals and remove the redundancy information.}
\begin{table}[!t]
\centering
\caption{The Architecture of CNN Network.}\label{tab:CNN}
{\scriptsize
\begin{tabular}{cccc}
       \Xhline{1.2pt}
          \multicolumn{1}{c|}{\footnotesize{Input Size}}  &\multicolumn{1}{c|}{\footnotesize{Layer Type}}  &\multicolumn{1}{c|}{\footnotesize{Stride}}  &\footnotesize{Kernel Number}  \\
        \hline
        \hline
           \footnotesize{$6 \times 128$}  &\footnotesize{conv $1\times9$}  &\footnotesize{2}   &\footnotesize{32} \\

           \footnotesize{$6\times64\times32$}  &\footnotesize{pool $1\times2$ } &\footnotesize{2 }  &\footnotesize{/ }\\

           \footnotesize{$6\times32\times32$}  &\footnotesize{conv $1\times3$ } &\footnotesize{1 }  &\footnotesize{64} \\

           \footnotesize{$6\times32\times64$}  &\footnotesize{conv $1\times3$ } &\footnotesize{1 }  &\footnotesize{128} \\

           \footnotesize{$6\times32\times128$ } &\footnotesize{pool $1\times2$}  &\footnotesize{2 }  &\footnotesize{/} \\

           \footnotesize{$6\times16\times128$}  &\footnotesize{conv $6\times1$ } &\footnotesize{1 }  &\footnotesize{128 }\\

           \footnotesize{$1\times16\times128$}  &\footnotesize{inner product } &\footnotesize{0}   &\footnotesize{/ }\\
        \Xhline{1.2pt}
\end{tabular}

}
\end{table}

\begin{figure}[!t]
\centering
\includegraphics[width=0.7\linewidth]{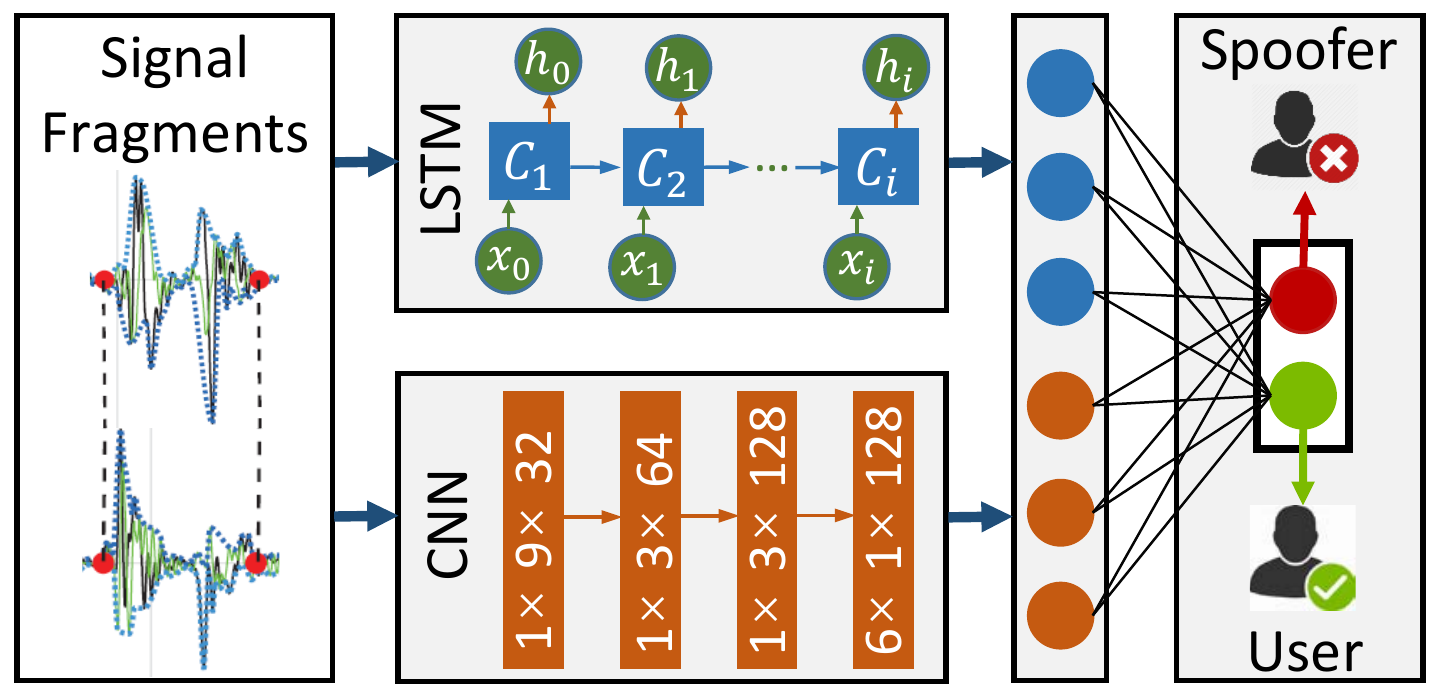}
\vspace{-2mm}
\caption{\red1{The architecture of the LSTM-CNN network.}}
\label{fig:lstmcnn}
\vspace{-3mm}
\end{figure}

\noindent \textbf{Lip Motion Detection.}\label{sec:Motion_Detection}
After \textit{Character Segmentation}, we can preliminarily determine whether each segmented signal fragment contains the information of lip motions by calculating the Signal to Noise Ratio (SNR). However, some signal fragments may be produced by the stubborn dynamic interference rather than ``physical'' lip motions.
\qi{Thus, we train a deep learning model to decide whether each signal fragment corresponds to the ``physical'' lip motions for reading a character in the passcode. The model training will not be impacted by signal features associated with the specific passcode. Therefore, we do not require retraining the model even when the user changes the passcode.}
Theoretically, the long short-term memory (LSTM) network can effectively extract temporal features, and the convolutional neural network (CNN) can well handle spatial features. We build a LSTM-CNN network to detect lip motions, which can effectively learn both   temporal and spatial features.

Each segmented fragment can be considered as a feature vector. Since the length of feature vectors that are fed into CNN must be consistent, FaceLip first re-samples all fragments into 128-dimensional feature vectors.
\red1{
Then each vector is fed into both LSTM  and CNN, as shown in Fig.~\ref{fig:lstmcnn}.
The LSTM used in FaceLip has two layers with 64 hidden neurons in each layer. The LSTM can be seen as a feature extractor with a 64-dimensional output $f_{LSTM}(x)$.
The CNN contains 4 convolution layers and 2 max-pooling layers. The feature of lip signals along the time-series is abstracted by the convolution kernel, and the feature map is down-sampled by the max-pooling. In Table~\ref{tab:CNN}, we present the architecture of our CNN network and the parameters of each layer. We denote the output of CNN as $f_{CNN}(x)$. The features $f_{LSTM}(x)$ and $f_{CNN}(x)$ are then catenated and fed into a fully-connected layer whose output is a two-dimensional vector, indicating whether the lip motion is valid.}
Since the number of characters in each passcode is known (\eg, 4) in FaceLip, \red1 {theoretically the user passes \textit{Motion Verification} only if all fragments (each fragment corresponds to ``physical'' lip motions of speaking a character) are detected. However, some signal fragments may be misjudged due to the low SNR. Based on our tests, we empirically set a threshold, \emph{i.e.}, if more than half of the signal fragments are detected, the user is supposed to pass \textit{Motion Verification}.
}


\begin{figure}[!t]
\centering
\includegraphics[width=0.8\linewidth]{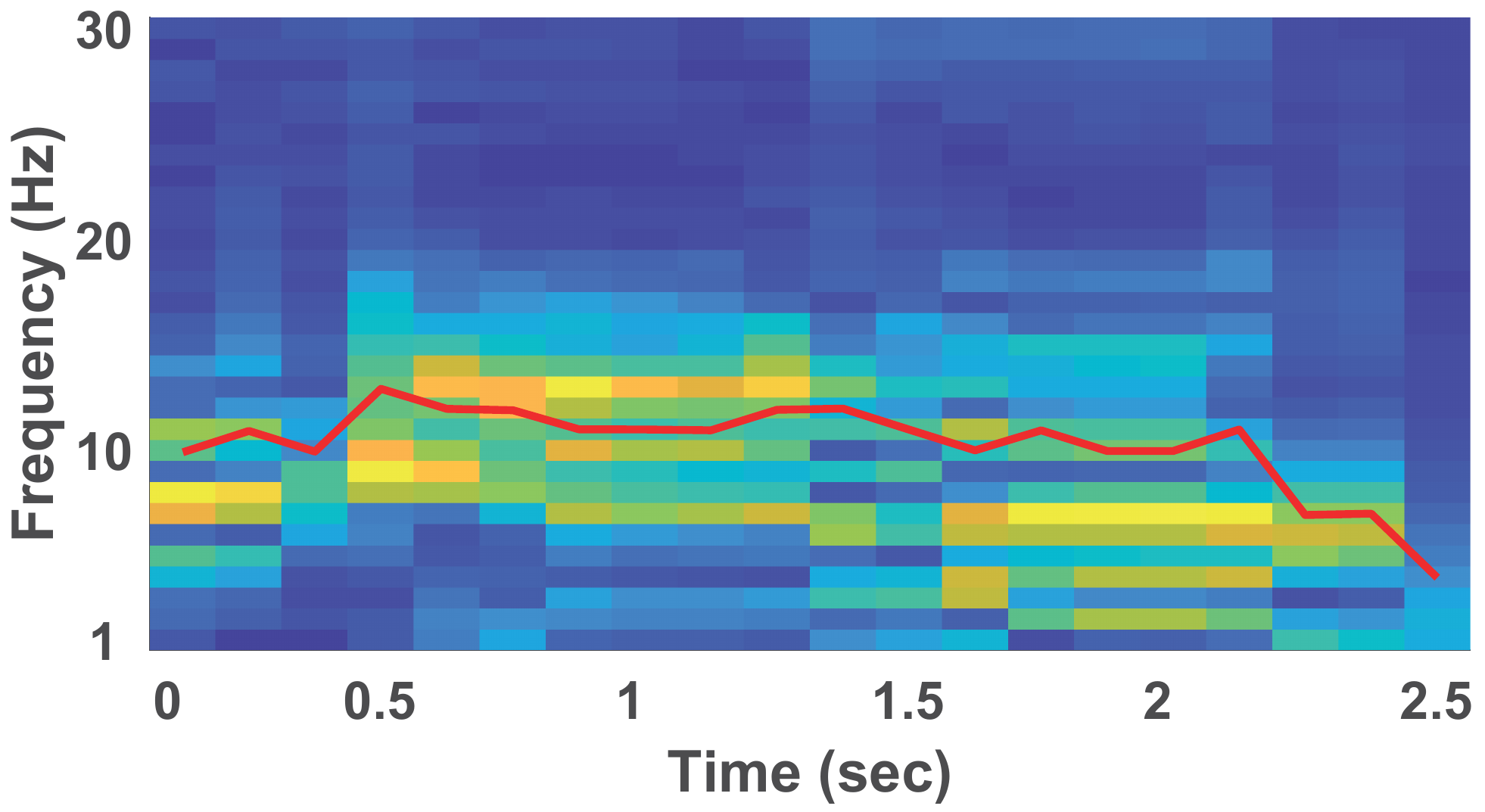}
\vspace{-2mm}
\caption{\red1{The energy-band time-frequency features.}}
\label{fig:feature}
\vspace{-3mm}
\end{figure}

\subsection{Consistency Verification}
\red1{
  Now we conduct consistency verification by comparing captured lip motions and that collected during the user registration phase after the lip motion is detected as ``valid'', which can effectively distinguish legitimate attempts from malicious ones. We divide the verification into two steps:  \textit{Feature Extraction} and \textit{Consistency Decision}.
}

\red1{
\noindent\textbf{Feature Extraction.} We first splice together all signal fragments corresponding to characters and convert the spliced signal from the time domain to the frequency domain by short-time Fourier transform (STFT) with a window size of 1000ms and an overlap size of 875ms. Fig.~\ref{fig:feature} shows a spectrum example of the spliced signals when a user speaks ``4M8R''. It shows that the energy distribution of Doppler shifts caused by lip motions varies over time. Inspired by ~\cite{zhang2017hearing}, we extract energy-band time-frequency features as follows. We first normalize energy values in the spectrum into the same range (\ie, $0 \sim 1$), and then calculate the accumulated energy values of all frequencies at each time point. Note that, to avoid the interference of noise signal, only the energy values at the range of $0.03 \sim 0.99$ will be considered. Finally, we find the centroid frequency where the accumulated energy values are the half of that of all frequencies over the time points and combine all the centroid frequencies of each time point together as the energy-band time-frequency features.
}

\noindent\textbf{Consistency Decision.}
\qi{FaceLip leverages the profile of each user to build a binary classifier for the user through Support Vector Machine (SVM). To achieve this, we construct a unique user profile for each legitimate user. The user's own lip motion features extracted during the user registration are set as positive samples of the profile. Note that, FaceLip uses lip motion features of other users as initial negative samples of the profile that are shipped with FaceLip. In each round of liveness detection, FaceLip takes the extracted lip motion features of a user as the input of the SVM binary classifier to determine whether the extracted features are consistent with the positive samples of the user profile. Due to the uniqueness of the user's lip motion~\cite{faraj2006motion}, it is impossible for an attacker to bypass the consistency verification by impersonating legitimate users. Note that, after each round of detection, either positive or negative samples of the profile will be updated with the lip motion features extracted from this input for the next round of liveness detection.}

\section{Security Analysis}\label{sec:securityanalysis}

In this section, \re2{
we first analyze the security properties of FaceLip by abstracting the liveness detection mechanism as a challenge-response based protocol. 
} Then, we analyze the robustness of FaceLip against wide-ranging sophisticated attacks.

\subsection{Security Properties of FaceLip}
FaceLip could be treated as a challenge-response based protocol. Its security is guaranteed by enabling random challenges, unforgeable acoustic signal responses, and effective response confirmations. The challenges in FaceLip are the $N$ randomly generated frequencies of the carrier signal. According to our threat model, the front-end device of the authentication system is well protected, and thus the attacker can not know the random frequencies in advance.
Note that, the number of different carrier frequencies in our design is around $2.3 \times 10^9$. Obviously, it is almost unlikely for the attacker to conjecture all frequencies correctly.
The responses in FaceLip are acoustic signals containing the information of lip motions. 
Since lip motion patterns of different individuals are unique and hard to imitate~\cite{faraj2006motion}, the responses corresponding to the challenges contain distinct characteristics of the individual and are very hard to forge. Thus, \textit{Motion Verification} together with the \textit{Consistency Verification} allows FaceLip to realize effective response confirmations and can defeat existing attacks. 

Specifically, our proposed liveness detection scheme is built upon unforgeable lip motions. It utilizes \textit{Motion Verification} and \textit{Consistency Verification} modules to enable secure liveness detection.
\textit{Motion Verification} detects whether the recorded audio contains ``physical'' lip motions, which ensures that the information of lip motions extracted from the acoustical signals are only produced from ``physical'' movements rather than ``visual'' movements. Therefore, this module can effectively resist attacks only contain ``visual'' lip motions. Sophisticated attacks also cannot cheat FaceLip by involving ``physical'' lip movements produced by the attacker since the uniqueness of each individual's lip motion pattern makes it unforgeable. \textit{Consistency Verification} abstracts the characteristics of each individual's lip motions through the time-frequency features and then feeds the characteristics into SVM to verify the consistency, which guarantees that the entire lip motions are generated from the real user.


\subsection{Security Against Sophisticated Attacks}\label{sec:various_attack}
\re2{
FaceLip enables a tie-breaking mechanism of lip motion-based liveness detection by leveraging unforgeable acoustic signals. Existing attacks can be classified into two categories.}

\re2{\noindent{\textbf{Attacks without Constructing Valid Lip Motions}.} The \reda3{MFF attacks (\ie, the image based attack and the video based attack),} use medias containing victims' faces to fool face authentication systems. The VR based attack collects images of the victim on social networks to reconstruct the 3D face and displays the face video on the screen of a VR device. 
These attacks only contain ``visual'' lip movements rather than ``physical'' movements. 
The ``visual'' lip motions cannot pass \textit{Motion Verification} in FaceLip, and thus the attacks will fail under FaceLip.}

\re2{\noindent{\textbf{Attacks with Constructing Valid Lip Motions}.} Lip motions contained in the acoustic signals can be generated by forging the signals or reflecting the audio via lips. We can further classify existing attacks into two categories, i.e., 2D attacks (\ie, replay attacks and synthesis attacks) and 3D dynamic attacks (\ie, 3D mask attacks, adversarial attacks and 3D projection attacks).}

\re2{Replay attacks record acoustic responses of previous detection processes and replay them to fool the current detection process. 
FaceLip chooses $N$ subcarriers and randomizes the frequency of each subcarrier. Theoretically, the hitting possibility of frequencies of all subcarriers in FaceLip is around $\frac{1} {2.3 \times 10^9}$, which is difficult for an attacker to evade FaceLip. Moreover, an attacker could extract the victim's lip motion features from the previous responses and embed them into the carrier signal of the current challenges to construct synthesis attacks, which cannot evade FaceLip either. 
First, the acoustic signals that are used to capture the lip motions in FaceLip are specially designed, which cannot be captured by an attacker to predict in advance. 
Second, it is hard to forge current acoustic responses in real time because it is impossible for the attacker to ensure the integrity of lip motion features under twice multiple path interferences in the environment.} 

The 3D mask attack is performed by wearing a custom 3D silicone mask to impersonate the victim. To reconstruct the 3D face structure of the victim, the attacker needs to use multiple high-quality pictures constructed from different angles of the victim. In addition, manufacturing such custom silicone masks is complicated and costly. 
Similarly, the adversarial attack impersonates the victim by designing unobtrusive perturbations (\eg, wearing eyeglasses) around the attacker's face according to the adversarial examples. 
\qi{Moreover, the 3D projection attack projects the  victim's reconstructed face videos on a 3D face model and then generates the responses on the
basis of the challenges.}
Fortunately, FaceLip can still defeat these attacks because the attacker cannot imitate the victim's lip motion patterns to bypass \textit{Consistency Verification} in FaceLip.

\begin{figure*}[!th]
\begin{minipage}[t]{0.327\linewidth}
\centering
\includegraphics[width=0.98\textwidth]{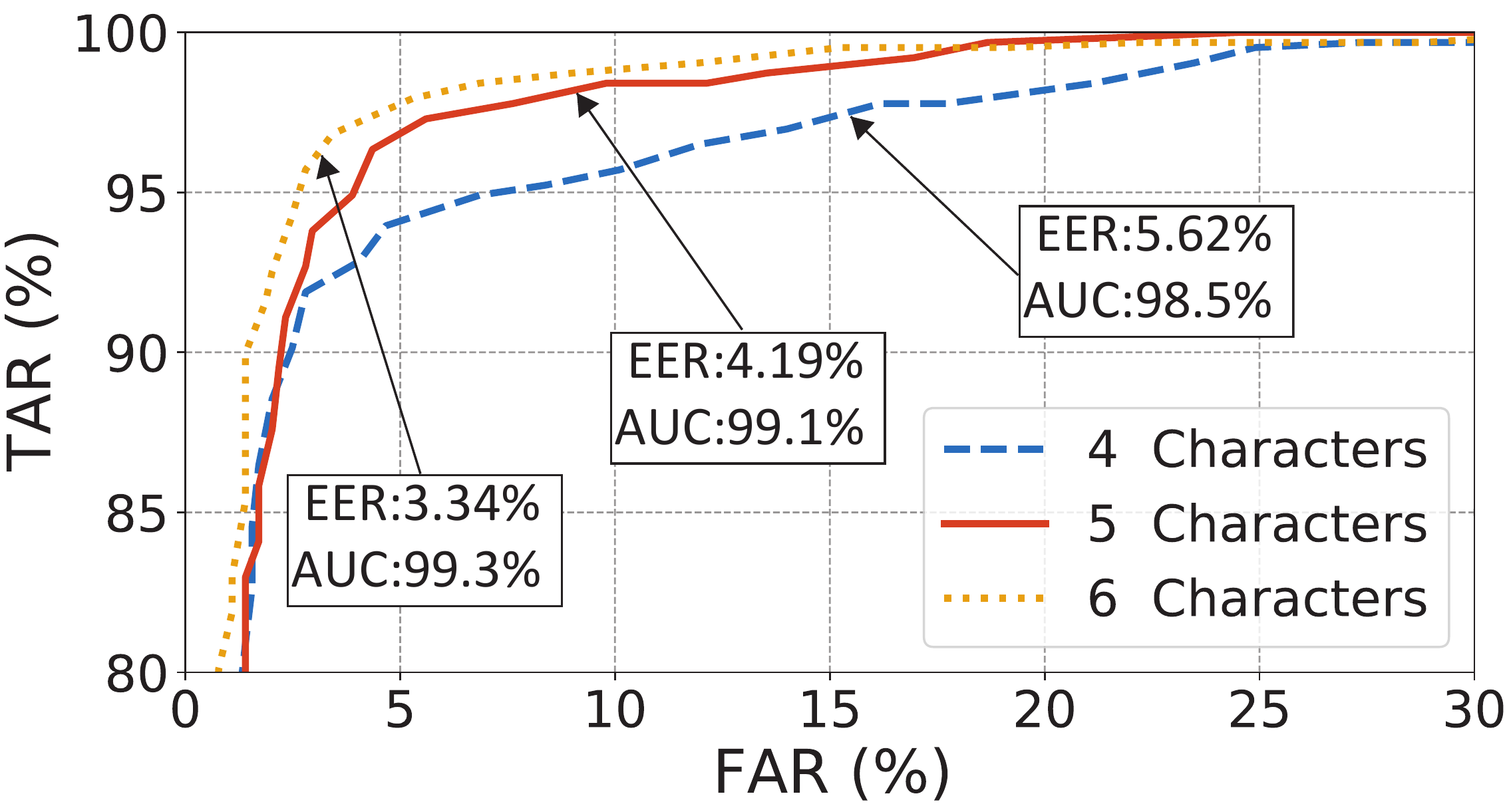}
\vspace{-2mm}
\caption{ROC curves of motion verification.}
\label{fig:liveness_ROC}
\end{minipage}
\begin{minipage}[t]{0.327\linewidth}
\centering
\includegraphics[width=0.98\textwidth]{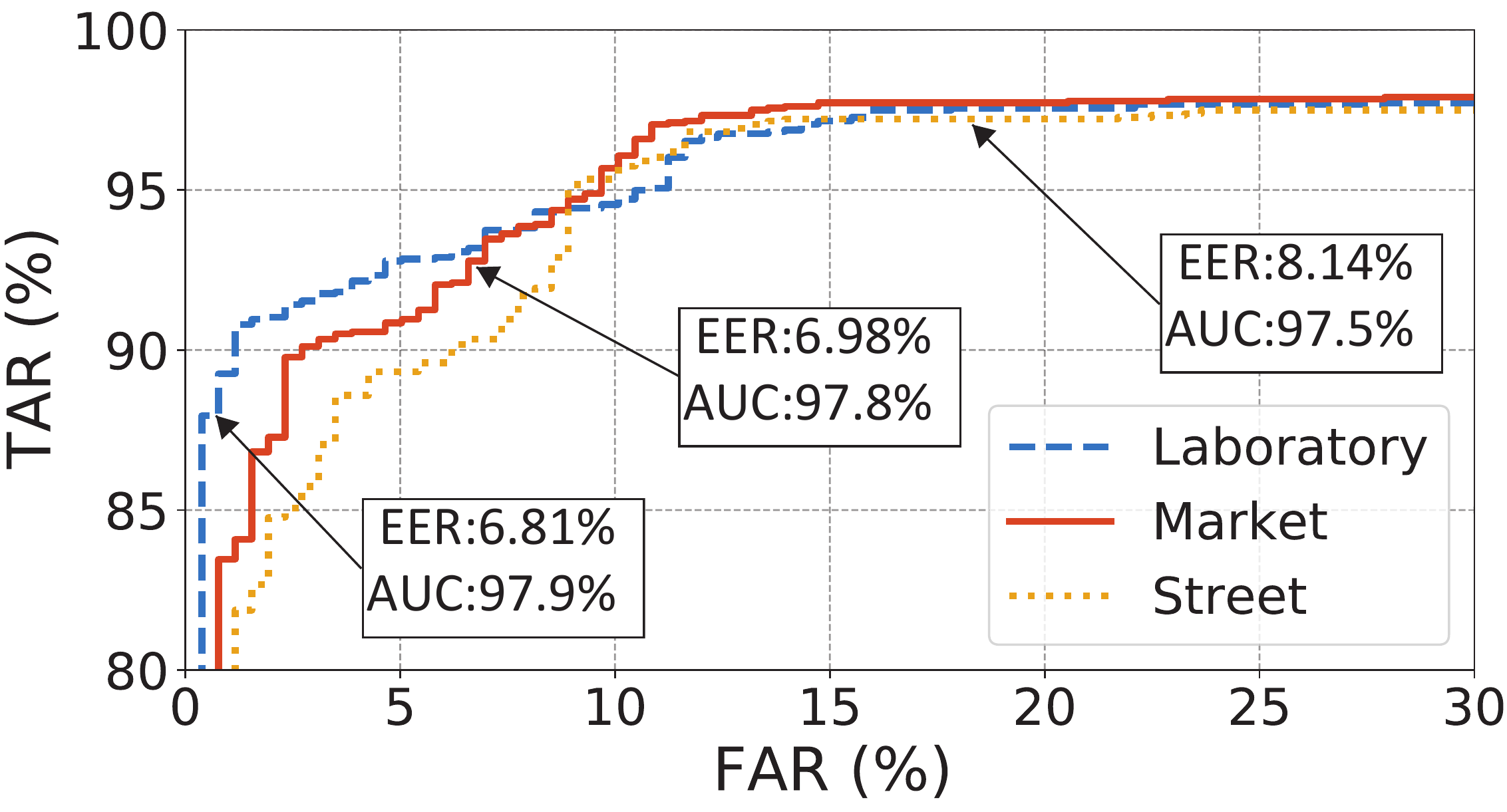}
\vspace{-2mm}
\caption{ROC curves of consistency verification.}
\label{fig:consistency_ROC}
\end{minipage}
\begin{minipage}[t]{0.327\linewidth}
\centering
\includegraphics[width=0.98\textwidth]{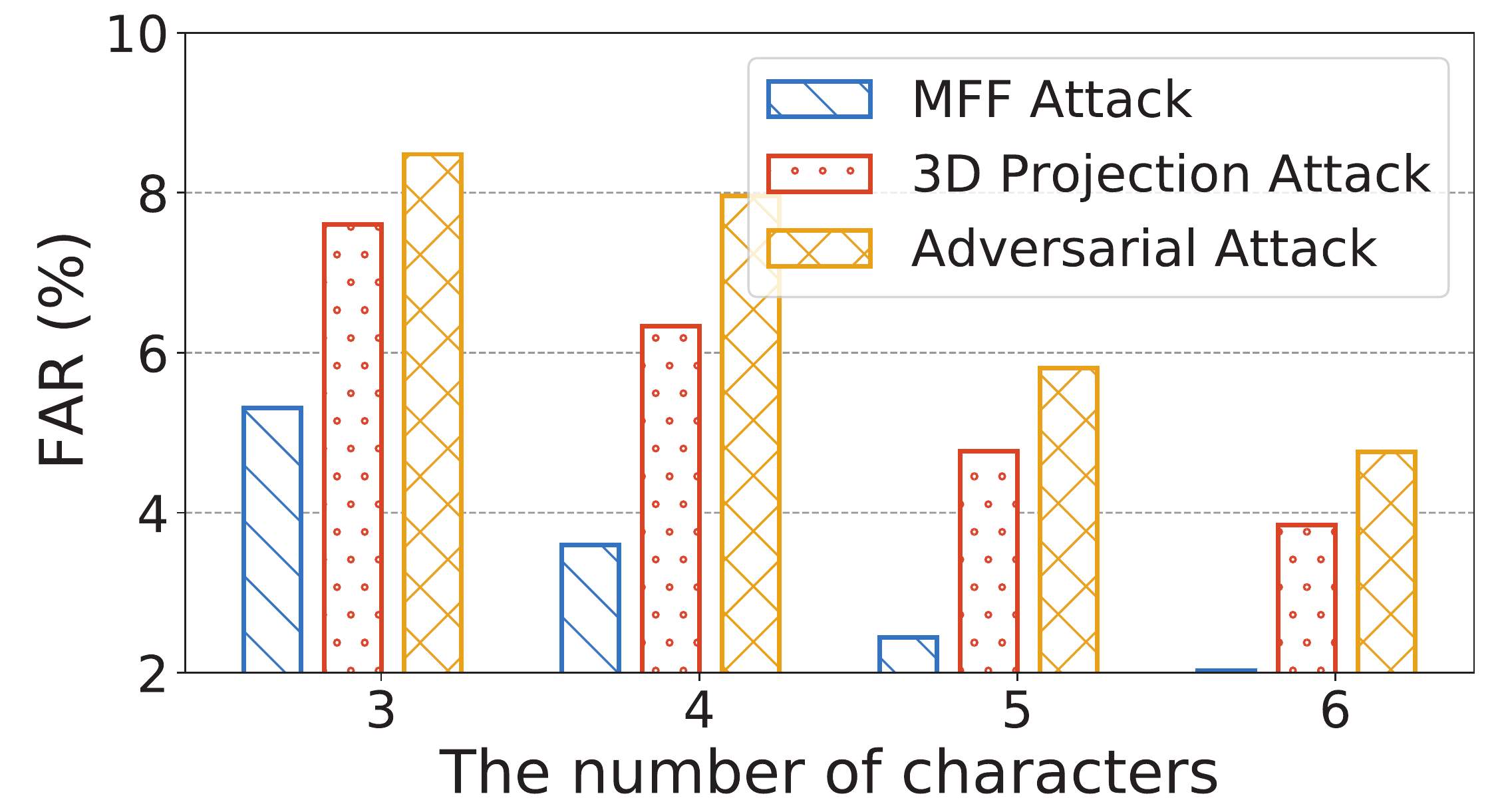}
\vspace{-2mm}
\caption{FAR under various attacks.}
\label{fig:various_attack}
\end{minipage}
\vspace{-3mm}
\end{figure*}

\section{Evaluation}\label{sec:evaluation}

In this section, we present the experimental results based on our FaceLip prototype built on off-the-shelf smartphones.

\subsection{Experimental Setup}

\noindent{\textbf{Implementation}.}
We implement a FaceLip prototype on off-the-shelf smartphones to evaluate its performance and effectiveness. We evaluate FaceLip on three different smartphone platforms: a SAMSUNG C9 Pro running Android 6.0.1, a Xiaomi Mi9 SE running Android 9, and a HUAWEI Honor7 running Android 6.0. The played acoustic signals are the superposition of multiple tones whose frequencies are in the range of $18\sim 21 $KHz. Then, the recorded audios are uploaded to the server for signal  analysis. 
The server is a PC with 3.1GHz CPU and 8GB memory. Note that there is more than one speaker and microphone in most mobile devices. However, most smartphones have the issue of hardware echo cancellation when under dual track recording, which can affect the recorded signals from different pairs of speaker-microphone. Therefore, for simplicity, we only use one microphone to record the acoustic signals reflected by the lips.

\noindent{\textbf{Data Collection}.}
Our experiments involve 44 participants aged from 18 to 39, including 27 males and 17 females. For each scenario, each volunteer speaks his/her passcode towards the smartphone during face liveness detection. 
In the user registration phase, the user chooses his/her passcode with different lengths.
Each participant is required to speak the passcode for 5 times \re2{under different settings (\eg, different distances, environments, passcode lengths, and phone models) at the enrollment phase to build the user-specific model and then perform liveness detection 5 times under different settings for testing. In total, the dataset used for evaluation includes 18,920 samples.} Note that all volunteers were informed and signed consents for our data collection. Actually, we only analyze the acoustic signals reflected by lips and store features extracted from the signals instead of the collected signals. We confirm that these features cannot be used to identify users.

\begin{figure*}[!th]
\begin{minipage}[t]{0.327\linewidth}
\centering
\includegraphics[width=0.98\textwidth]{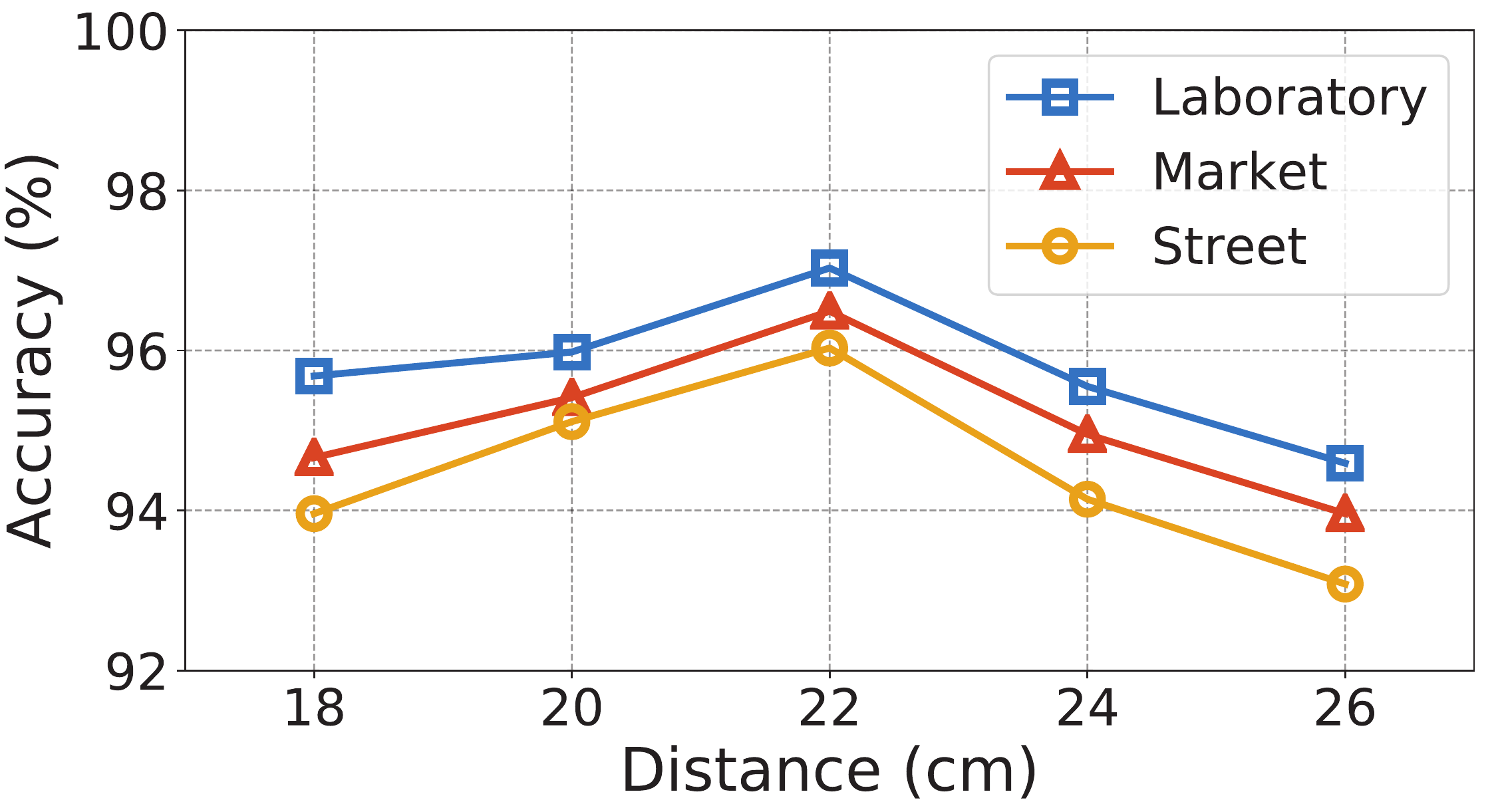}
\vspace{-2mm}
\caption{Impact of distance and environment.}
\label{fig:liveness_distance}
\end{minipage}
\begin{minipage}[t]{0.327\textwidth}
\centering
\includegraphics[width=0.98\columnwidth]{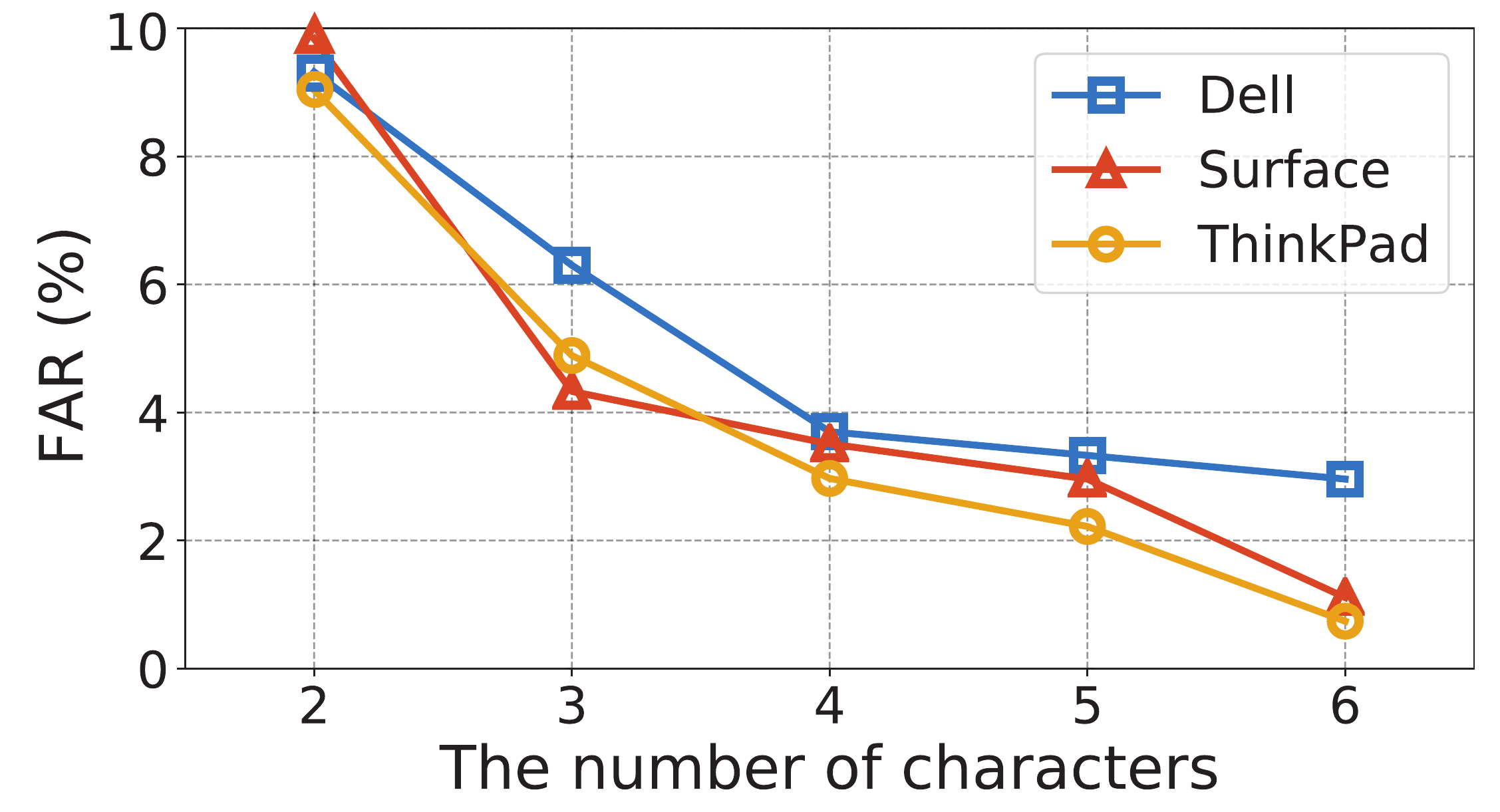}
\vspace{-2mm}
\caption{Impact of passcode length.}
\label{fig:liveness_password}
\end{minipage}
\begin{minipage}[t]{0.327\textwidth}
\centering
\includegraphics[width=0.98\columnwidth]{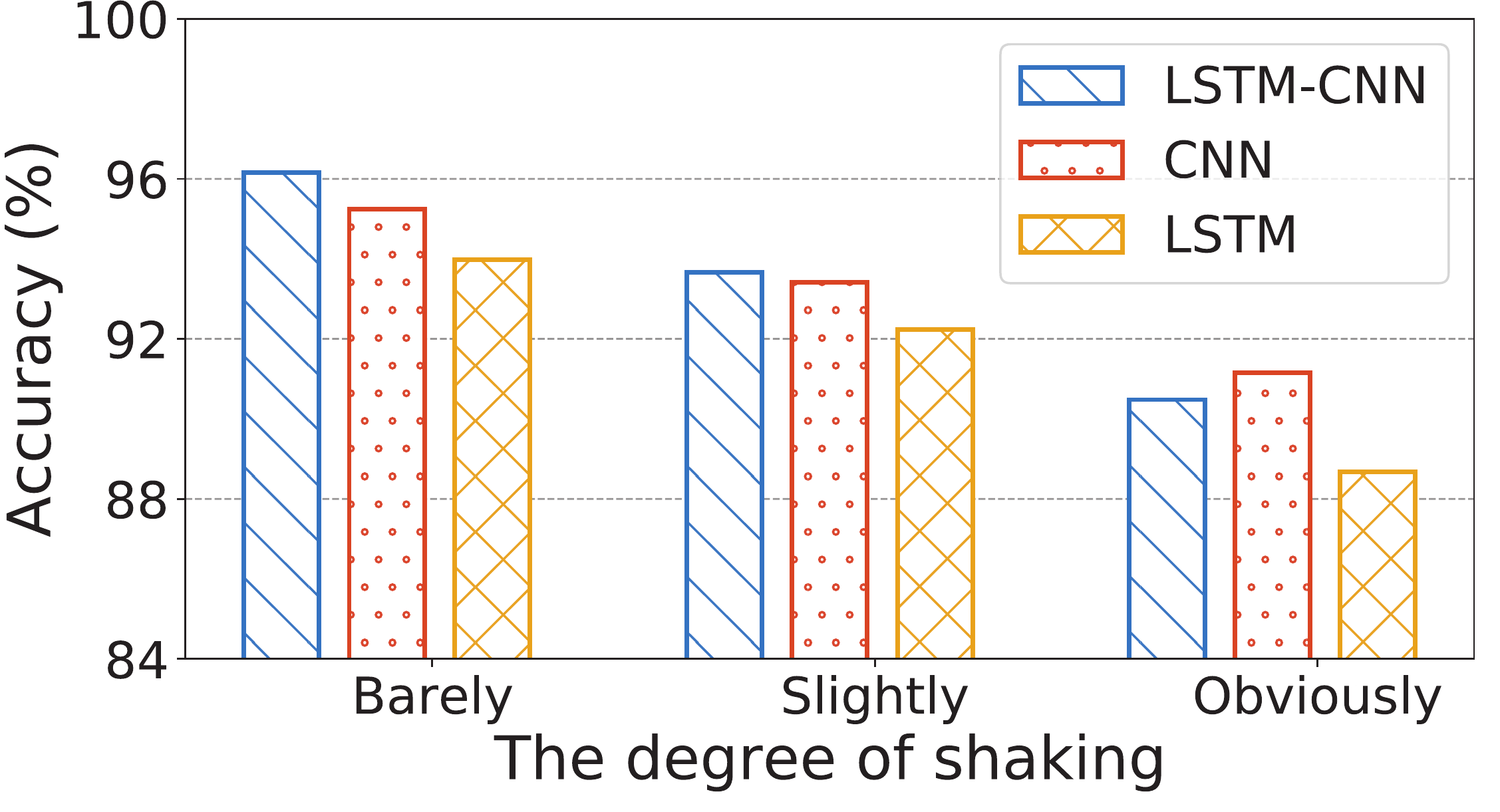}
\vspace{-2mm}
\caption{Impact of shaking and classifier.}
\label{fig:liveness_shake}
\end{minipage}
\vspace{-3mm}
\end{figure*}

\begin{table}[!t]
\centering
\caption{The parameters of different screens.}\label{tab:screen}
{\scriptsize
\begin{tabular}{c||ccc}
       \Xhline{1.2pt}
       \footnotesize{ Screen } &\multicolumn{1}{c}{\footnotesize{Resolution}}  &\multicolumn{1}{c}{\footnotesize{Size}}& \footnotesize{Pixel density } \\
        \hline
        \hline
        \footnotesize{Dell S2340}  & \footnotesize{$1920 \times 1080 $ } & \footnotesize{23 inches} &\footnotesize{95.8 ppi} \\

        \footnotesize{ Surface Pro 4} & \footnotesize{$2736 \times 1824 $} & \footnotesize{12.3 inches} &\footnotesize{267.3 ppi} \\

        \footnotesize{ThinkPad X1}  & \footnotesize{$2560 \times 1440 $} & \footnotesize{14 inches} & \footnotesize{210 ppi}\\

       \Xhline{1.2pt}
\end{tabular}

}
\vspace{-3mm}
\end{table}

\noindent{\textbf{Default Setting}.}
The generated acoustic signals are in the range of $18 \sim 21$KHz with 3 random frequencies since the multiple feature vectors obtained from different frequencies can be combined to mitigate the multipath effect, and the random frequencies can resist the audio replay attack.
The sampling rate of the microphone is 48KHz. We used the top speaker and the bottom microphone to generate and record acoustic signals. Our experiments are conducted under three different environments: a laboratory (quiet), a market (slightly noisy), and a street (noisy). The length of the passcode ranges from $2 \sim 6$ characters. The laboratory is the default environment, and we choose 4 characters for the passcode unless claimed otherwise. In our experiments, by default, we used SAMSUNG C9 Pro as the evaluation platform, and let the participant sit on the chair and perform liveness detection when the smartphone is fixed at a distance from the phone's microphone to the user's lips of 18cm.

\noindent{\textbf{Attacks}.}
We evaluate FaceLip under MFF attacks, the 3D projection attack, and the adversarial attack. In MFF attacks, we record the victim's facial videos to build the adversarial videos and then display the videos on three different screens (\ie, Dell, Surface, and ThinkPad). The parameters are shown in Table~\ref{tab:screen}.
For the 3D projection attack, we project the adversarial videos on the 3D silicone face model as described in Section~\ref{sec:projection attack} and then use FaceLip to perform face liveness detection towards the projected face model. For the adversarial attack, we specifically start the impersonate attack in~\cite{sharif2016accessorize}. The volunteers are asked to wear specialized glasses (which can help to pass the face authentication) and try to imitate the victim's lip motions as much as possible. We validate each type of attack against all 44 participants with 660 times of attempts and measure the average value.

\noindent{\textbf{Metrics}.}
\re2{
To evaluate the performance of FaceLip, we use the following metrics. True Accept Rate (TAR) is the possibility that FaceLip identifies the legitimate users correctly. False Accept Rate (FAR) is the probability that FaceLip identifies the attacker as a legitimate user.
The Receiver Operating Characteristic (ROC) curve describes the relationship between TAR and FAR under a varying classification threshold. The area under the ROC curve (AUC) is the probability that prediction scores of
legitimate users are higher than attackers. 
False Reject Rate (FRR) is the probability  that FaceLip identifies a legitimate user as the attacker.
Equal Error Rate (EER) defines the rate when FRR equals FAR.
Accuracy means the probability that Facelip distinguishes between the legitimate users and the attackers correctly. 
}

\subsection{Overall Performance}
\noindent{\textbf{ROC Curves}.}
We use the ROC curve to show the overall performance of FaceLip under different passcode lengths and environments. Fig.~\ref{fig:liveness_ROC} and~\ref{fig:consistency_ROC} demonstrate the results of the \textit{Motion Verification} and \textit{Consistency Verification} modules,  respectively. As shown, FaceLip can well distinguish between legitimate users and MFF attackers when the passcode length is more than 3. \re2{The ERR and AUC of \textit{Motion Verification} are 3.34\% and 99.3\% for the lengths of 6 characters, respectively, which shows good performance of FaceLip. Besides, FaceLip can also well distinguish authenticated users and 3D dynamic attackers in different environments. The ERR and AUC of \textit{Consistency Verification} are 6.18\% and 97.9\% in a laboratory, and it validates that FaceLip works well under the real environments.}


\noindent{\textbf{FAR Against Various Attacks}.}
To evaluate FaceLip's robustness to various attacks, we test the FAR of FaceLip against the MFF attack, the 3D projection attack, and the adversarial attack with different passcode lengths. As shown in Fig.~\ref{fig:various_attack}, \re2{the FAR can only reach 3.6\% under the MFF attack, 6.33\% under the 3D projection attack, and 7.96\% under the adversarial attack when the passcode length is 4.}
\qi{The reason is that the MFF attack only contains ``visual'' lip movements rather than ``physical'' movements and cannot pass \textit{Motion Verification} in FaceLip, and the 3D projection attack and the adversarial attack cannot imitate the victim's unique lip motion patterns to bypass \textit{Consistency Verification} in FaceLip.}
Therefore, FaceLip can throttle these attacks effectively.

\subsection{Motion Verification}
\noindent{\textbf{Experiment Details}.} We use Tensorflow, an open-source deep learning framework, to construct and train the CNN and LSTM model used in lip motion detection. The model is trained at a learning rate of 0.001 and a max iteration of 500. The positive samples are built from the raw acoustic signals (live person) recorded by FaceLip and the negative samples are built from the acoustic signals in MFF attacks because the \textit{Motion Verification} module mainly aims to defeat \reda3{the typical presentation attacks,~\eg, MFF attacks.} The training dataset includes 7,754 positive samples and 5,147 negative samples, and the testing dataset includes 9,940 positive samples and 6,435 negative samples.


\noindent{\textbf{Impact of Distance and Environment}.}
In this experiment, we investigate the influence of liveness detection distance and environment on the accuracy of \textit{Motion Verification}. The distance, which is carefully measured from the phone's microphone to the user's lips, varies from 18cm to 26cm on the horizon. 
Fig.~\ref{fig:liveness_distance} demonstrates the accuracy under different distances and environments.
We find that the accuracy increases first and then decreases when the distance increases. FaceLip has the best performance (97.03\% accuracy in a laboratory) when the distance is 22cm. This is because there is strong signal interference from pop noises~\cite{wang2019voicepop} when the mouth is too close to the phone, while acoustic signals decay rapidly as the propagation distance increases. When the distance is larger than 22cm, it leads to an obvious loss of lip motion information.
In addition, we can observe that the ambient noise in different environments does not influence the performance of \textit{Motion Verification} obviously. This is mainly due to the fact that the generated acoustic signals are in the range of $18 \sim 21$KHz, and ambient noise becomes insignificant under these frequencies.

\noindent{\textbf{Impact of Passcode Length}.}
This experiment evaluates the influence of passcode length on the FAR of \textit{Motion Verification}. Fig.~\ref{fig:liveness_password} shows the FAR with different passcodes with the lengths varied from 2 to 6 characters with different attack devices. The results show that the FAR decreases gradually as the passcode length increases. Passcodes with longer lengths carry more amount of lip motion information for the liveness detection of individuals. Generally,  the higher the screen resolution of the attack device, the higher the success rate of MFF attacks. Therefore, we generate the negative samples using three devices with different screen resolutions in Table~\ref{tab:screen}.
An interesting observation is that a higher screen resolution will not increase the success rate of MFF attacks against FaceLip. It is because we use the acoustic signals rather than visual signals to conduct \textit{Motion Verification}.

\begin{figure}[!t]
\centering
\includegraphics[width=0.7\columnwidth]{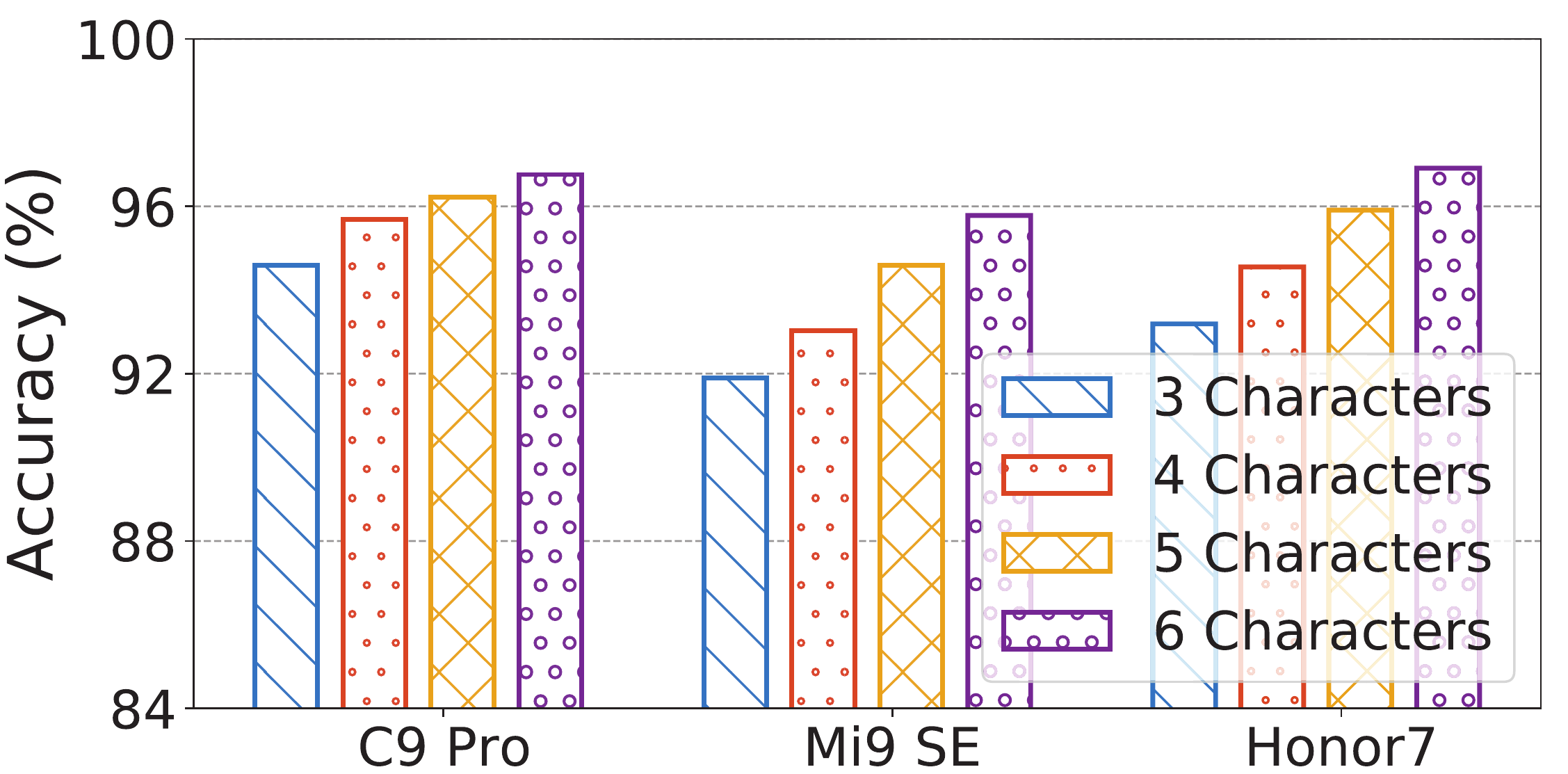}
\vspace{-2mm}
\caption{Impact of phone model.}
\label{fig:liveness_smartphone}
\vspace{-3mm}
\end{figure}

\begin{figure*}[!th]
\begin{minipage}[t]{0.327\linewidth}
\centering
\includegraphics[width=0.98\textwidth]{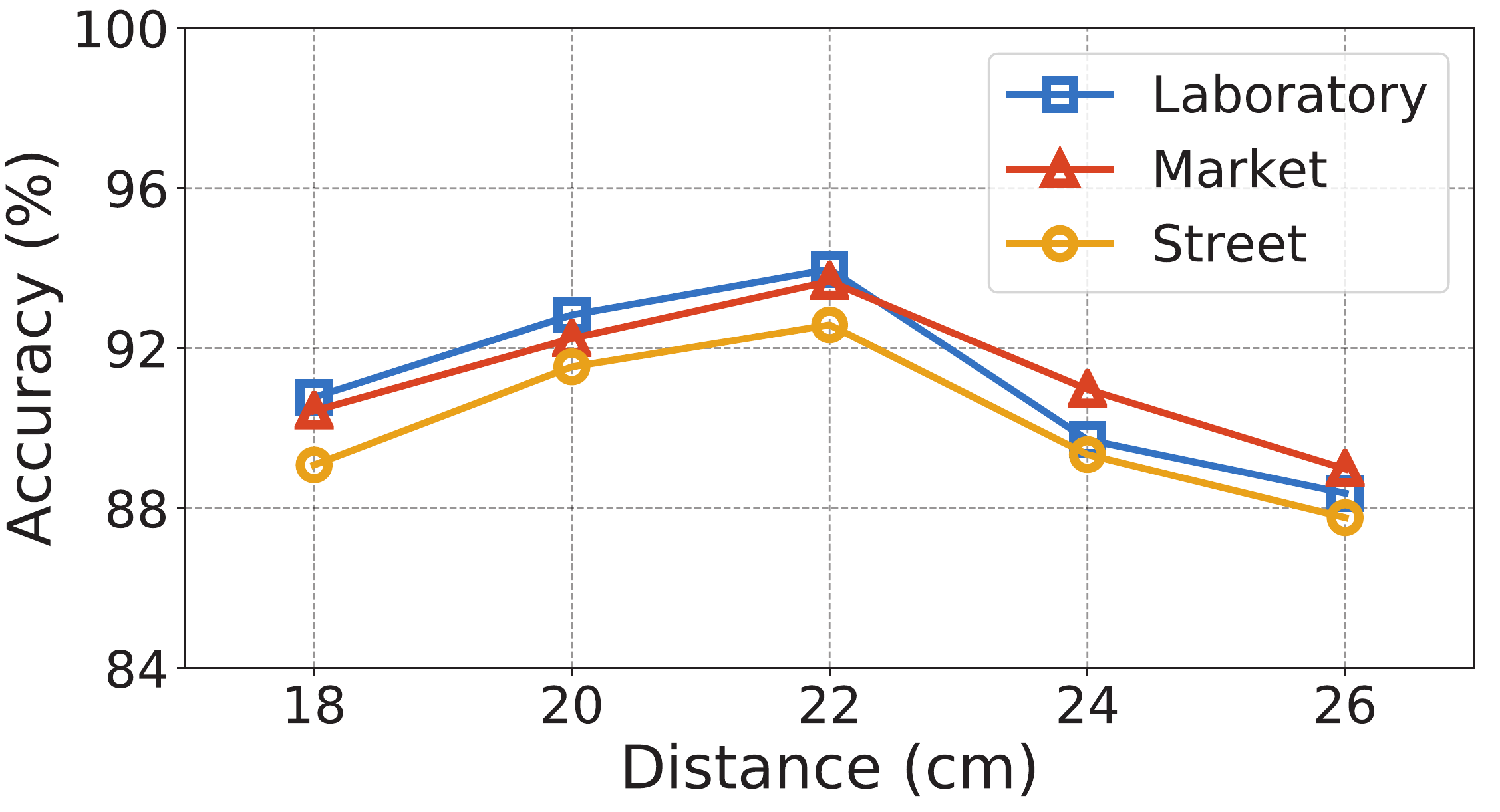}
\vspace{-2mm}
\caption{ Impact of distance and environment.}
\label{fig:consistency_distance}
\end{minipage}
\begin{minipage}[t]{0.327\textwidth}
\centering
\includegraphics[width=0.98\columnwidth]{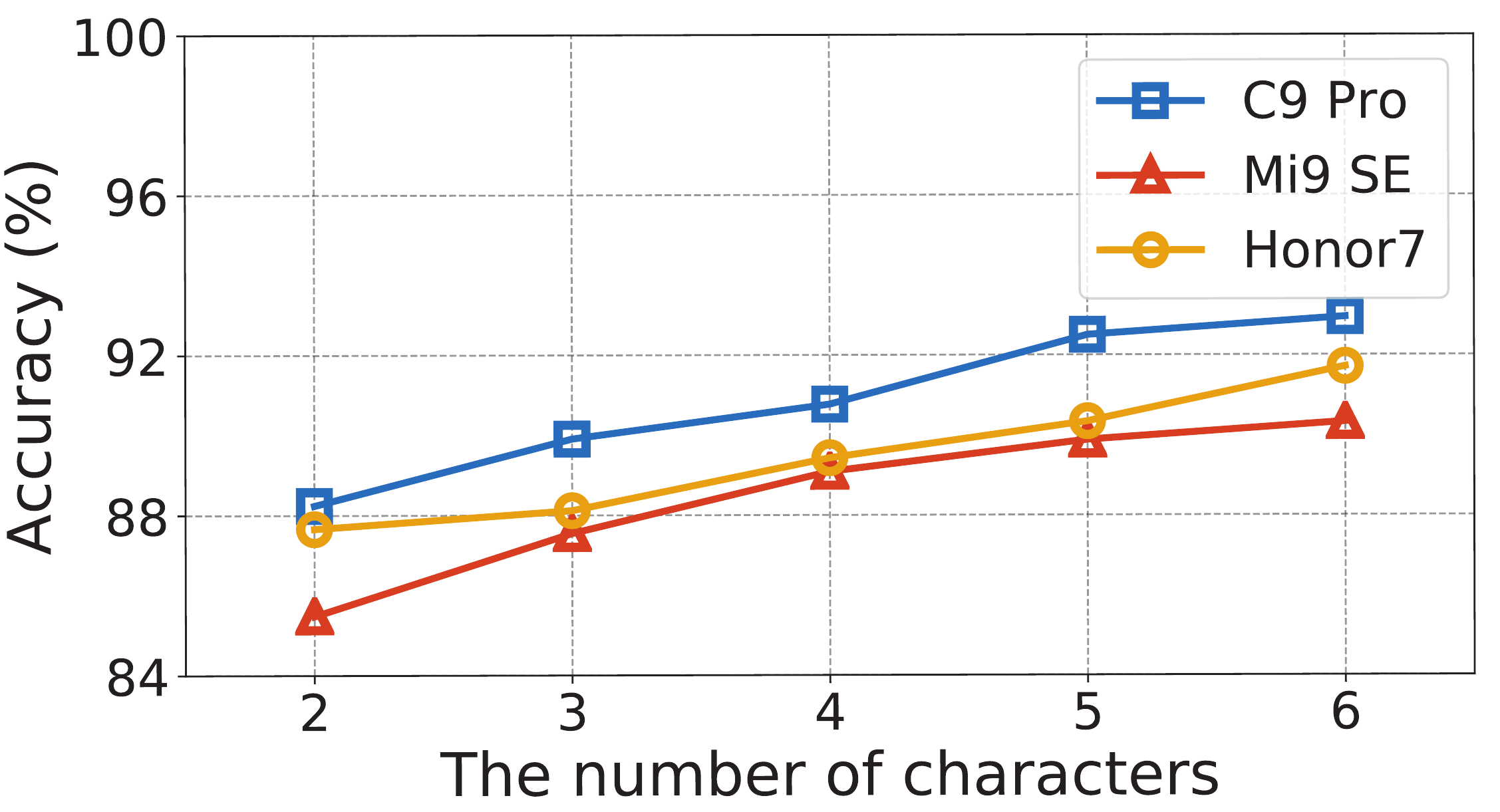}
\vspace{-2mm}
\caption{ Impact of passcode length and phone.}
\label{fig:consistency_password}
\end{minipage}
\begin{minipage}[t]{0.327\textwidth}
\centering
\includegraphics[width=0.98\columnwidth]{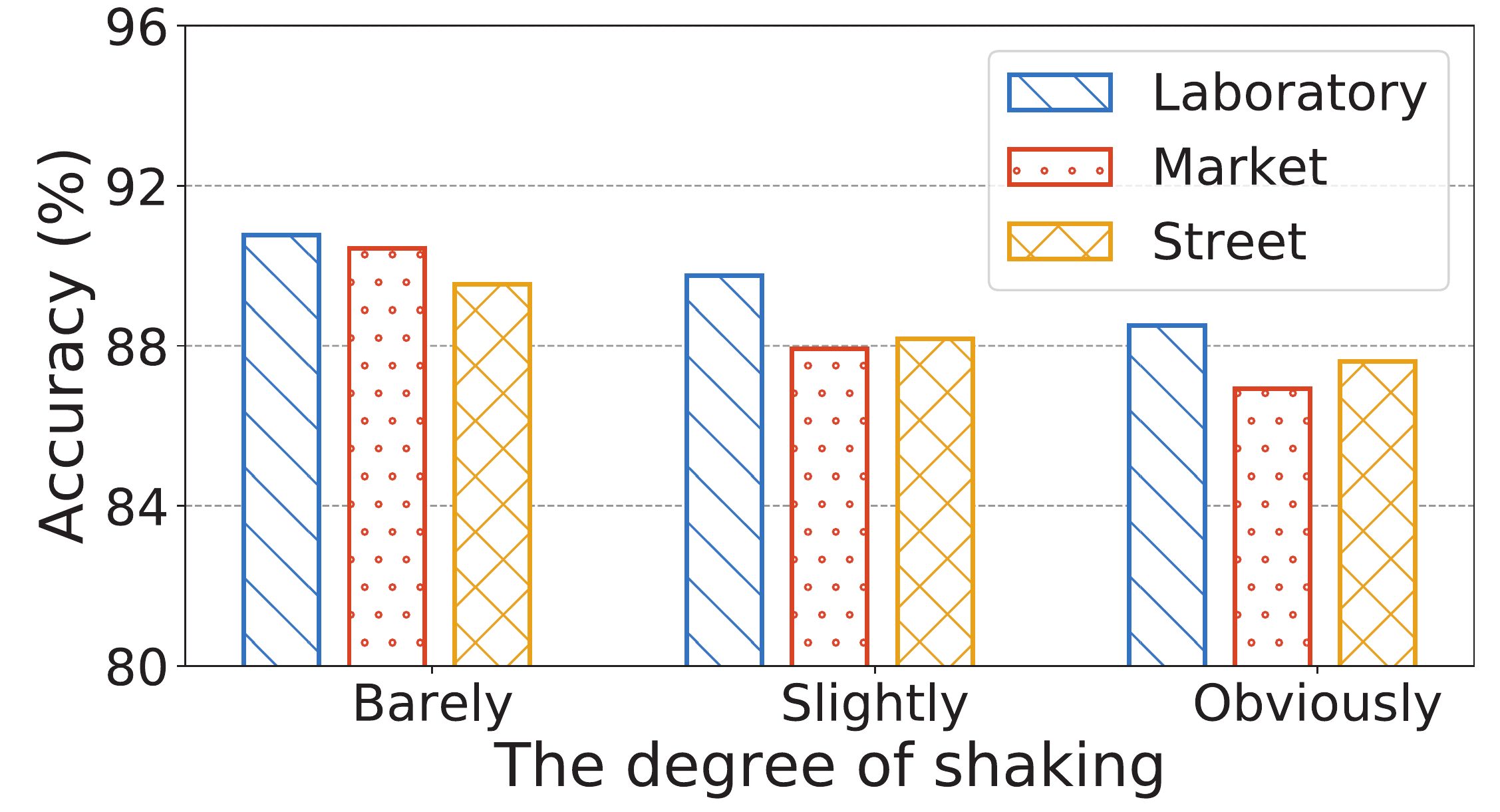}
\vspace{-2mm}
\caption{Impact of shaking.}
\label{fig:consistency_shake}
\end{minipage}
\vspace{-3mm}
\end{figure*}

\noindent{\textbf{Impact of Shaking and Classifier}.}
We further study the impact of shaking and classifier on the accuracy of \textit{Motion Verification}. Considering that the smartphone may have different degrees of shaking when the user holds the phone to perform face liveness detection, we evaluate the accuracy under three degrees of shaking: i) barely shaking, where the phone is fixed on a phone holder, ii) slightly shaking, where the phone is held by the user normally, and iii) obviously shaking, where the user holds the phone and shakes it intentionally. As shown in Fig.~\ref{fig:liveness_shake}, \re2{FaceLip performs the best with the accuracy of 95.15\% when the phone is shaken barely and has the lowest accuracy when the phone is shaken significantly.} When the smartphone is fixed on a phone holder (shaking barely), the dynamic interference is much less than that of shaking obviously. Note that the accuracy with slight shaking is only a bit lower than shaking barely. Thus, we can know that FaceLip is robust to slight disturbances of the body movement. Overall, we can see that the
LSTM-CNN network classifier performs better than a single CNN and a single LSTM network classifier. \qi{Note that, CNN alone is more appropriate for phones that are held with intentional shaking, while LSTM-CNN is more suitable for normal phone usages. Thus, we choose the LSTM-CNN network classifier in \textit{Motion Verification}.} 

\noindent{\textbf{Impact of Phone Model}.}
To study the effectiveness of FaceLip on different phone models, we installed FaceLip on three off-the-shelf smartphones: a SAMSUNG C9 Pro, a Xiaomi Mi9 SE, and a HUAWEI Honor7. The results are presented in Fig.~\ref{fig:liveness_smartphone}. As shown, the performance of C9 Pro is slightly better than these of Mi9 SE and Honor7. We infer that the acoustic sensors and signal processing module in SAMSUNG are better than the other two platforms, \reda3{thus the played and received acoustic signals suffer less distortion.} Besides, passcodes with longer length provide stronger protection. FaceLip achieves 97.14\% accuracy with a passcode of 6 characters.

\subsection{Consistency Verification}

\noindent{\textbf{Experiment Details}.}
To evaluate the effectiveness of consistency verification, we mainly test FaceLip's ability to \re2{distinguish legitimate attempts from malicious ones by analyzing the acoustic signals reflected by the lip motions.} We use cubicSVM implemented by Matlab to verify the performance of the lip motion detection. The SVM is trained with the kernel function in the polynomial function, while the penalty coefficient Box-Constraint is set to 1.
The positive samples are built from the raw acoustic signals from the same benign person, and the negative samples are built from the acoustic signals in adversarial attacks because the \textit{Consistency Verification} module mainly aims to defeat 3D dynamic attacks. A unique SVM model is constructed for each person, and we have 430 positive samples and 430 negative samples for evaluation.



\noindent{\textbf{Experimental Results}.}
We present the results of consistency verification in terms of the impact of distance, environment, passcode length, phone model, and shaking in this experiment.  
Fig.~\ref{fig:consistency_distance} shows the accuracy of consistency verification under different distances and environments. Similarly, the accuracy first increases and then decreases when the distance increases. The accuracy is always over 88\%, which means that consistency verification works well within our suggested distance for liveness detection.  Different environments also do not affect the performance of consistency verification obviously.
The impact of passcode length and phone model is shown in Fig.~\ref{fig:consistency_password}. A longer passcode enables stronger protection, and FaceLip performs well on different phone platforms. FaceLip achieves 92.95\% accuracy with a passcode of 6 characters. The degrees of shaking also affect the performance, as shown in Fig.~\ref{fig:consistency_shake}.
However, shaking slightly does not cause a significant drop in accuracy. Overall, the results demonstrate the robustness of the \textit{Consistency Verification} module in Facelip.

\section{Discussion}

\noindent{\textbf{Usability of FaceLip}.}
\qi{FaceLip utilizes unforgeable lip motion patterns encoded with passcodes built upon well-designed acoustic signals to ensure a strong security guarantee for liveness detection. The user is required to speak the passcode composed of 4$\sim$6 characters during the liveness detection. Note that, passcodes used in FaceLip are different from PINs that are explicitly input by users and vulnerable to various attacks, \eg, shoulder surfing attacks and smudge attacks. Existing defenses, \eg, rtCaptcha~\cite{uzun2018rtcaptcha}, and real production services 
~\cite{Tencent} leverage dynamic codes answered by users, which are extracted in the captured videos, to perform liveness detection. However, they cannot defeat sophisticated presentation attacks, \eg,
3D dynamic attacks.
Fortunately, we can implement similar dynamic passcode in FaceLip to eliminate remembering passcode, which can improve the user experience and be an interesting topic for future work.
}

\noindent{\textbf{Passcode Setup}.}
\qi{FaceLip aims to utilize unforgeable lip motion patterns encoded with passcodes to enable a secure face liveness detection scheme, which is not impacted by diverse populations, \eg, people with various skin-colors. }
We suggest users choose passcodes involving more obvious lip motions during the passcode setup. A passcode normally contains a series of syllables. The upper and lower lips will form particular shapes when producing different syllables during the speech, and the obvious lip shapes that significantly vary among different individuals can distinguish different users better. For instance, the passcode containing the syllable ``tu:'' is more suitable than ``$\alpha$:r'' because the syllable ``tu:'' generates more obvious lip motions than ``$\alpha$:r''.

\re2 {
\noindent{\textbf{Real Deployment with Various Phones}.}
FaceLip can be easily deployed in various devices with a camera, a speaker, and a microphone. We implement a FaceLip prototype as an app and install it on various off-the-shelf smartphones. It only requires the permissions approved by users to access the camera and microphone. We do not need to re-configure the smartphone systems or add any additional hardware. It is certain that our system can be deployed locally on devices, \eg, the smartphone. However, 
with the development of cloud services, it is not necessary to deploy our system locally, which can process the images and audios more efficiently. Thus, we suggest deploying FaceLip in a typical client-server architecture. Only \textit{Audio Capturing} is set up in the front-end device (\ie, smartphone), while \textit{Motion Verification} and \textit{Consistency Verification} are set up in the server. The captured audio is uploaded to the server for signal analysis in real time.
}

\re2 {
\noindent{\textbf{Retraining LSTM-CNN Networks}.}
Normally, FaceLip does not require retraining the LSTM-CNN network. We utilize the LSTM-CNN network to decide whether the signal fragment is corresponding to the ``physical" lip motions of reading a passcode. It will not be impacted by the signal features associated with the specific passcode. Thus, if a user wants to change the passcode, the user only needs to change the passcode with the trained model without retraining the LSTM-CNN network.
}

\section{Conclusion}
This paper studied the security of existing liveness detection methods in face authentication systems and found none of them can effectively defend against existing attacks, in particular sophisticated 3D dynamic attacks.
Particularly, we developed a new 3D projection attack that easily circumvents popular commercial facial recognition services with rather low costs and high universality. To defeat these attacks, we proposed FaceLip, which enables a tie-breaking mechanism of lip motion based liveness detection by leveraging unforgeable acoustic signals and ensures a strong security guarantee for face authentication systems. We implemented a FaceLip prototype on various off-the-shelf smartphones. Extensive experiments validate the effectiveness and robustness of FaceLip with various practical settings.


\bibliographystyle{IEEEtran}
\bibliography{FaceLip}

\end{document}